\documentclass[12pt,leqno]{article}
\usepackage[left=1in,right=1in,top=1in,bottom=1in]{geometry}
\usepackage{amssymb,amsmath,amsthm}
\usepackage{amsfonts,bbm}
\usepackage{blindtext}
\usepackage{booktabs}
\usepackage{bm}
\usepackage{caption}
\usepackage{comment}
\usepackage{enumerate}
\usepackage{float}
\usepackage{ragged2e}
\usepackage{graphicx}
\usepackage{bbm}
\graphicspath{{figures/}}
\usepackage{indentfirst}
\usepackage{mathtools}
\usepackage{mathrsfs}
\usepackage{multirow}
\usepackage{upgreek}
\usepackage[utf8]{inputenc}
\usepackage{xcolor}
\usepackage[bibstyle=numeric, citestyle=authoryear, natbib=true,doi=false, url=true, backend=biber, maxbibnames=6, maxcitenames=4, uniquelist=false, uniquename=false, sorting=nyt]{biblatex}
\renewbibmacro{in:}{}

\usepackage{array}
\usepackage{rotating}
\usepackage{setspace}
\usepackage{subcaption}
\usepackage{tikz}
\usepackage{xurl}
\usepackage[psdextra,hypertexnames=false]{hyperref}
\usepackage{bookmark}
\usepackage{cleveref}
\crefname{enumi}{}{}
\usepackage{dcolumn}
\usepackage{float}
\usepackage{longtable}
\usepackage{threeparttable}
\renewcommand{\arraystretch}{1}


\newtheorem{theorem}{Theorem}

\newtheorem{lemma}{Lemma}

\newtheorem{remark}{Remark}

\newtheorem{condition}{Condition}

\crefname{figure}{Figure}{Figures}
\crefname{algorithm}{Algorithm}{Algorithms}
\crefname{assumption}{assumption}{assumptions}
\crefname{lemma}{Lemma}{Lemmata}
\crefname{table}{Table}{Tables}
\crefname{condition}{Condition}{Conditions}

\numberwithin{theorem}{section}
\numberwithin{corollary}{section}
\numberwithin{stheorem}{section}
\numberwithin{lemma}{section}
\numberwithin{result}{section}
\numberwithin{definition}{section}
\numberwithin{assumption}{section}
\numberwithin{remark}{section}
\numberwithin{proposition}{section}
\numberwithin{condition}{section}
\numberwithin{equation}{section}

\newcommand{\indicator}[1]{\mathbbm{1}\{#1\}}
\newcommand{\indicatorbig}[1]{\mathbbm{1}\big\{#1\big\}}
\newcommand{\indicatorBig}[1]{\mathbbm{1}\Big\{#1\Big\}}
\newcommand{\indicatorBigg}[1]{\mathbbm{1}\Bigg\{#1\Bigg\}}

\newcommand{\E}{\mathbb{E}}
\newcommand{\Eps}{\mathcal{E}}
\newcommand{\Prob}{\mathbb{P}}
\newcommand{\R}{\mathbb{R}}
\newcommand{\X}{\mathcal{X}}
\newcommand{\x}{\bm{x}}
\newcommand{\xbf}{\mathbf{x}}
\newcommand{\xrm}{\mathrm{x}}
\newcommand{\y}{\bm{y}}

\newcommand{\V}{\mathbb{V}}
\newcommand{\dmet}{\mathrm{d}}
\newcommand{\Op}{\mathcal{O}_p}

\newfloat{sfigure}{tbp}{sfig}
\floatstyle{ruled}
\newfloat{algorithm}{tbp}{loa}
\floatname{algorithm}{Algorithm}
\floatstyle{plain}
\DeclareNameAlias{default}{family-given/given-family}

\begin{document}
    \title{A Grid-Rate Condition for Valid Uniform Inference}
	\date{\today }
	\author{Emmanuel Selorm Tsyawo\footnote{email: estsyawo@gmail.com, Department of Economics, Finance and Legal Studies, Culverhouse College of Business, University of Alabama}}
	\maketitle
	\thispagestyle{empty}
	\begin{abstract}\noindent
Conducting uniform inference on a continuous functional $F: \X \to \R$ involves specifying $L_n^d$ nodes on $\X \subset \R^d$ for estimation and the construction of confidence bands. While asymptotically valid inference requires $L_n$ to increase with $n$, existing fixed-$L$ rules of thumb and heuristic data-driven approaches lack formal justification. This paper shows that, for functions within a Donsker class, the simple grid-growth condition \(r_n^{1/4}/L_n\to0\), denoted \(L_n=\omega(r_n^{1/4})\), is sufficient for valid inference on twice continuously differentiable functions whose estimators satisfy \(r_n^{1/2}(\widehat F-F)=\Op(1)\). 
 
		\vspace{1.5cm}
		
		
		\vspace{0.25cm}
		
		
	\end{abstract}

\newpage
\setcounter{page}{1}

\section{Introduction}
The practice of uniform inference on continuous functionals \(F(\x)\), \(\x \in \X \subset \mathbb{R}^d\), \(d<\infty\), such as distribution functions \citep{chernozhukov-fernandez-melly-2013inference} or quantile functions \citep{callaway-li-2019}, via confidence bands, typically relies on estimating the underlying \emph{continuous} object only at a \emph{finite} collection of grid points \( \X^{L} \subset \X \). At first sight, this creates a tension between the inherently continuous nature of the target functional and its discrete implementation. In practice, however, this tension is reconciled by interpolating across the grid points. The estimates and associated ``uniform'' confidence bands are displayed as continuous curves obtained by linearly connecting the nodes. As emphasised by \citet[Sec.~2.4]{epperson-2013-introduction}, ``almost all graphs produced by computers are actually the result of something called \emph{piecewise linear interpolation}...'' The point is not merely cosmetic: once the reported object is an interpolated curve or surface, interpolation becomes part of the inferential procedure rather than an after-the-fact plotting convention. Exploiting linear interpolation ensures the estimators of the linear interpolant $F_L(\cdot)$ (the estimand of $F(\cdot)$ \emph{de facto}) and its confidence bands retain the continuity property of $F$ over $\X$. 

This paper derives a simple grid-growth rate that requires $L_n$ to grow strictly faster than $r_n^{1/4}$ (denoted $L_n = \omega(r_n^{1/4})$), under which the resulting confidence bands are asymptotically valid.\footnote{$r_n=n$ for functionals whose estimators are pointwise consistent at the usual $\sqrt{n}$ parametric rate.} The analysis thus provides a theoretically grounded rule for choosing \(L_n\). While \citet{imai-2025-accuracy} works in a non-Donsker non-parametric setting and delivers a case-specific coverage bound to be inverted by the practitioner, the Donsker framework in the current paper, which covers semiparametric estimators such as quantile and distribution regression, admits a clean rate-transparent condition: $L_n = \omega(n^{1/4})$, with $r_n = n$ pinned down by the $\sqrt{n}$-rate.

The paper makes three contributions. First, it formalises the interpolated confidence band as the reported inferential object, rather than treating interpolation as a purely graphical post-processing step. Second, it derives a deterministic interpolation-error bound that yields the grid-growth condition \(L_n=\omega(r_n^{1/4})\) for twice continuously differentiable targets. Third, it shows that, under a Donsker representation and a standard bootstrap approximation, the interpolated sup-\(t\) bands have the same asymptotic uniform coverage as the infeasible continuous-index benchmark.

Obtaining a grid-growth rate for asymptotically valid uniform inference matters because a critical trade-off emerges. On the one hand, a sparse grid with slow-growing or fixed $L_n$ is computationally cheap but may fail to capture the function's variability between points, leading to confidence bands that are asymptotically invalid, i.e., their true coverage probability falls below the nominal level. On the other hand, a dense grid (fast-growing $L_n^d$, especially when $d>1$) is more likely to yield valid uniform coverage but can be computationally prohibitive and may make the finite-sample supremum more sensitive to noisy pointwise estimates, scale estimates, or shape irregularities such as quantile crossing, even when the latter can be repaired by rearrangement.

The existing literature often acknowledges that the number of nodes $L_n$ ought to grow at a rate sufficient for the discrete maximum to converge to its continuous supremum, but the corresponding grid-growth rule is often left implicit. For example, \citet[Remark 3.1]{chernozhukov-fernandez-melly-2013inference} recommends $L_n = \omega(n^{1/2}) $ and \citet[Sect. 4.1]{kim-wooldridge-2024-difference} suggests a fixed-L default $L_n=99$. This paper provides a clear guideline for practitioners by linking computational grid choices to asymptotic validity. The argument connects the continuity of the target object with the discrete nature of its implementation through the interpolation error induced by replacing \(F\) with \(F_L\).

The Monte Carlo evidence in \Cref{subsec:mc_evidence} illustrates the downside of overly sparse grids: fixed grids perform reasonably for flat or very smooth targets but can under-cover when the target function has curvature that interpolation must resolve, whereas the proposed rate-guided rule performs comparably to denser power-rule grids while using fewer nodes.

The rest of the paper proceeds as follows. \Cref{sec:interp_framework} formalises the linear and multilinear interpolation framework and derives the deterministic interpolation-error bound. \Cref{sec:inference} shows that, under the grid-growth condition, the interpolated empirical process inherits the weak limit of the underlying process and the resulting interpolated confidence bands have valid uniform coverage. The section also records the computational algorithm and discusses Hadamard differentiable transformations. \Cref{subsec:mc_evidence} reports a Monte Carlo study of the proposed grid rules, and \Cref{sect:concl} concludes.

\paragraph{Notation:} For positive sequences \(a_n\) and \(b_n\), \(a_n=\omega(b_n)\) means \(b_n/a_n\to0\) as \(n\to\infty\). Whenever necessary, quantities such as \(L\) are indexed by \(n\), e.g., \(L_n\), to emphasise their dependence on the sample size. Let \(C\in(0,\infty)\) denote a generic constant whose value may vary across occurrences. For real numbers \(a\) and \(b\), define \(a \wedge b := \min\{a,b\}\) and \(a \vee b := \max\{a,b\}\). For a random variable \(V\), let \(Q_V(\tau)\) denote its population \(\tau\)-quantile, while \(\widehat Q_V(\tau)\) denotes the corresponding empirical or conditional-bootstrap quantile. The index set \(\X\) is a compact subset of \(\R^d\) equipped with the metric \(\dmet(\x,\y):=\lVert\x-\y\rVert_\infty\). For a vector \(\x\in\R^d\), \(\displaystyle \lVert\x\rVert_\infty:=\max_{1\leq k\leq d}|x_k|\); for a bounded function \(G:\X\to\R\), \(\displaystyle \lVert G\rVert_\infty:=\sup_{\x\in\X}|G(\x)|\).

\section{Interpolation Framework}\label{sec:interp_framework}
For the sake of clarity, the interpolation framework is first presented for the univariate case $d=1$ before being generalised to the $\R^d$-indexed empirical process.
\subsection[One-dimensional interpolation on R]{One-dimensional interpolation on $\R^d$}
It is useful first to discuss the construction in the special case $d=1$. For tractability, consider an equally spaced grid of $L$ nodes over the domain $\X := [\underline{\xrm}, \, \bar{\xrm}]$:
\[
\X^L := \{ \xrm_1, \xrm_2, \ldots, \xrm_L\}, \quad \text{where} \quad \xrm_{\ell+1} - \xrm_\ell = \varepsilon \text{ for } \ell = 1, \ldots, L-1,
\]
with mesh
\[
\varepsilon = \frac{\bar{\xrm} - \underline{\xrm}}{L-1}.
\]
The endpoints are tied at $\xrm_1 = \underline{\xrm}$ and $\xrm_L = \bar{\xrm}$, regardless of the choice of $L$. Denote the estimator of $F(\cdot)$ at $x \in \X $ by $ \widehat{F}(x) $. The inferential goal is uniform inference on $F(\cdot)$ over $\X$, while estimation is only feasible on a discrete set of nodes, say $\X^L \subset \X $. Linear interpolation converts the discrete set of grid-point estimates to a continuous estimator on $\X$.

For each \(x\in\X\), let \(\ell_L(x)\in\{1,\ldots,L-1\}\) denote the index of the grid interval containing \(x\):
\begin{equation}\label{eqn:ell_index_uni}
	\ell_L(x)
	:=
	\sum_{\ell=1}^{L-1}
	\ell\,\indicator{x\in[\xrm_\ell,\xrm_{\ell+1})},
	\qquad
	\ell_L(\xrm_L):=L-1.
\end{equation}
Thus \(x\in[\xrm_{\ell_L(x)},\xrm_{\ell_L(x)+1}]\), with the right-endpoint convention above. Define the local interpolation coordinate
\[
\delta_L(x):=
\frac{x-\xrm_{\ell_L(x)}}{\varepsilon}.
\]
For $x \in \X $, the linear interpolant of $F(x)$ is
\begin{align*}
   F_L(x)
   &:=
   \underbrace{\big(1-\delta_L(x)\big)}_{=:w_L(x;0)}F(\xrm_{\ell_L(x)})
   +
   \underbrace{\delta_L(x)}_{=:w_L(x;1)}F(\xrm_{\ell_L(x)+1}).
\end{align*}
Let $\widehat{F}_L(\cdot)$ be the same linear interpolant built from the estimated values $ \big\{ \widehat F(\xrm_1), \ldots, \widehat F(\xrm_L) \big\} $. In commonly displayed results, $\widehat{F}_L(x)$, $x\in\X$, is the reported continuous estimator of $ F(x)$, $x\in \X $. Linearly interpolating $F$ with $F_L$ on $\X$ incurs an interpolation error $R_n(\cdot)$:
\begin{equation}\label{eqn:rho_f_expand}
    \begin{split}
  \widehat{Z}_n^L(x):&= r_n^{1/2}\big(\widehat{F}_L - F\big)(x) = \underbrace{r_n^{1/2}\big(\widehat{F}_L-F_L\big)(x)}_{=:Z_n^L(x)}
+ \underbrace{r_n^{1/2}\big(F_L-F\big)(x)}_{=: R_n(x)} \\
&= Z_n^L(x) + R_n(x),
\end{split}
\end{equation} where \((r_n)_{n\ge 1}\) is a non-stochastic effective-rate sequence; \(x \in \X\), the centred and scaled estimator \(Z_n(x) := r_n^{1/2}\big(\widehat F - F\big)(x), \, x\in\X\) converges in distribution as \(n \to \infty\); \(\displaystyle Z_n^L(x):= w_L(x;0)Z_n(\xrm_{\ell_L(x)}) + w_L(x;1)Z_n(\xrm_{\ell_L(x)+1}) \) is the interpolated empirical process; and \(R_n(x)\) is the interpolation error.\footnote{\(r_n=n\) in the usual \(\sqrt n\)-regular case and possibly different under alternative effective sample sizes.} The multivariate process is discussed next.

\subsection[Multilinear interpolation on Rd]{Multilinear interpolation on $\R^d$}

The graphical motivation for linear interpolation is most transparent in the univariate case, where displayed curves are obtained by connecting adjacent grid-point estimates. In higher dimensions, the graphical interpretation is less literal, but the statistical role of interpolation is unchanged: it provides an in-sample continuous extension of grid-point estimates over \(\X\), and its deterministic interpolation error is what governs the grid-rate condition.

A tractable interpolation device in the multivariate setting is multilinear interpolation, with linear interpolation in the preceding subsection as the case $d=1$. Multilinear interpolation extends linear, bilinear, and trilinear interpolation to $d$-dimensional axis-aligned hyper-rectangles:
\[
\X := \prod_{k=1}^d [\underline{\xrm}_k,\, \bar{\xrm}_k],
\]
i.e., the Cartesian product of coordinate-wise intervals in $\R^d$. For multilinear interpolation on hyper-rectangles, $F$ is evaluated at the $2^d$ vertices of a cell
\[
\mathfrak{C}_{\bm{\ell}} = \prod_{k=1}^d [\xrm_{\ell_k, k},\, \xrm_{\ell_k+1, k}],
\]
indexed by $\bm{\ell} = (\ell_1,\ldots,\ell_d) \in \{1,\ldots,L-1\}^d $. On each axis $k\in \{1,\ldots,d\}$, there are $L$ equally spaced nodes with mesh
\(
\varepsilon_k = \frac{\bar{\xrm}_{k} - \underline{\xrm}_{k}}{L-1}, \qquad k = 1,\dots,d,
\)
where $\xrm_{1,k}$ and $\xrm_{L,k}$ are set to the endpoints $\underline{\xrm}_k$ and $\bar{\xrm}_k$, respectively. Then, observe that the index set $ \X $ is the union of all cells:
\[
\X = \bigcup_{\bm{\ell} \in \{1,\ldots,L-1\}^d} \mathfrak{C}_{\bm{\ell}} = \prod_{k=1}^d [\underline{\xrm}_k,\, \bar{\xrm}_k].
\]

For \(\x=(x_1,\ldots,x_d)\in\mathcal X\), let
\(
\bm\ell_L(\x)
=
\big(\ell_{L,1}(x_1),\ldots,\ell_{L,d}(x_d)\big)
\)
denote the cell index of \(\x\). This is the coordinate-wise analogue of the univariate interval index in \eqref{eqn:ell_index_uni}, with
\[ 
\ell_{L,k}(x_k)
:=
\sum_{\ell=1}^{L-1}
\ell\,\mathbf 1\{x_k\in[\xrm_{\ell,k},\xrm_{\ell+1,k})\}, \quad \ell_{L,k}(\xrm_{L,k})=L-1.
\]
Equivalently, \(\bm\ell_L(\x)\) is the unique multi-index selected by the coordinate-wise right-endpoint convention, and the selected closed cell satisfies
\(
\x\in\mathfrak C_{\bm\ell_L(\x)}.
\)
Define the local interpolation coordinates
\[
\delta_{L,k}(\x)
:=
\frac{x_k-\xrm_{\ell_{L,k}(x_k),k}}
{\varepsilon_k},
\qquad
\varepsilon_k
=
\xrm_{\ell_{L,k}(x_k)+1,k}
-
\xrm_{\ell_{L,k}(x_k),k}.
\]
The \(d\)-linear interpolant is then given by
\begin{align*}
F_L(\x)
&=
\sum_{\bm{\iota}\in\{0,1\}^d}
F\big(\xbf_L(\x;\bm{\iota})\big)
\prod_{k=1}^d
\big(1-\delta_{L,k}(\x)\big)^{1-\iota_k}
\delta_{L,k}(\x)^{\iota_k}\\
&=
\sum_{\bm{\iota}\in\{0,1\}^d}
F\big(\xbf_L(\x;\bm{\iota})\big)
\underbrace{\prod_{k=1}^d
\Big(
(1-\iota_k)
+
(2\iota_k-1)\delta_{L,k}(\x)
\Big)}_{w_L(\x;\bm{\iota})}\\
&=
\sum_{\bm{\iota}\in\{0,1\}^d}
w_L(\x;\bm{\iota}) F\big(\xbf_L(\x;\bm{\iota})\big),
\end{align*}
where
\( \displaystyle 
\xbf_L(\x;\bm{\iota})
=
\Big(
\xrm_{\ell_{L,1}(x_1)+\iota_1,1},
\dots,
\xrm_{\ell_{L,d}(x_d)+\iota_d,d}
\Big),
\quad
\bm{\iota}=(\iota_1,\ldots,\iota_d)\in\{0,1\}^d
\)
enumerates the $2^d$ vertices of the cell containing $\x$; see, e.g., \cite{weiser-1988-note}. The interpolant \(F_L(\x)\) is a convex combination of the vertex values
\(
F\big(\xbf_L(\x;\bm{\iota})\big), \ \bm{\iota}\in\{0,1\}^d
\),
since the weights
\(
w_L(\x;\bm{\iota})
\),
\(\bm{\iota}\in\{0,1\}^d\), are non-negative and satisfy
\( \displaystyle 
\sum_{\bm{\iota}\in\{0,1\}^d}
w_L(\x;\bm{\iota})
=
1.
\)
Moreover, for any $\xbf\in\X^L$, \(
F_L\big(\xbf_L(\xbf;\bm{\iota})\big)
=
F\big(\xbf_L(\xbf;\bm{\iota})\big)
\quad
\text{for all } \bm{\iota}\in\{0,1\}^d
\).

\subsection{Interpolation error}

For the multivariate interpolant defined above, define the resulting interpolation error by
\[
R_n(\x):=r_n^{1/2}\big(F_L-F\big)(\x),
\qquad \x\in\X.
\]
The following conditions on $\X$ and $F$ are useful in characterising this deterministic interpolation error.
\begin{condition}\label{cond:cond_compact_X}
	$ \X=\prod_{k=1}^d[\underline{\xrm}_k,\bar{\xrm}_k] $ is a compact rectangle in $\R^d$. Let
	\( \displaystyle 
	S_{\X}:=\max_{1\leq k \leq d}(\bar{\xrm}_k-\underline{\xrm}_k)
	\)
	denote its maximum coordinate span.
\end{condition}

\begin{condition}\label{cond:f_second_deriv_bound}
	$F$ is twice continuously differentiable, and
	\[
	M_F:=\sup_{\x\in\X}\max_{1\leq k\leq d}
	\left|\partial^2F(\x)/\partial x_k^2\right|
	<\infty .
	\]
\end{condition}

\noindent The following result characterises the interpolation error and the growth rate of $L$ in $n$ needed to make it asymptotically negligible.

\begin{lemma}\label{lem:interp_error}
	If \Cref{cond:f_second_deriv_bound,cond:cond_compact_X} hold, then
	\[
	\sup_{\x \in \X } |R_n(\x)|
	\leq
	\frac{dM_F S_{\X}^2}{8}
	\left(\frac{r_n^{1/4}}{L-1}\right)^2.
	\]
	Consequently, \( \displaystyle \sup_{\x \in \X } |R_n(\x)| = o(1) \) whenever $L_n = \omega(r_n^{1/4}) $.
\end{lemma}

\noindent The rate condition follows directly from the scaled interpolation error in \Cref{lem:interp_error}:
\[
\sup_{\x\in\X}|R_n(\x)|
=
r_n^{1/2}\|F_L-F\|_\infty
=
\mathcal{O}\left(\left(\frac{r_n^{1/4}}{L}\right)^2\right),
\]
with constant \(dM_F S_{\X}^2/8\) in the bound above. Hence, the interpolation error is asymptotically negligible whenever \(r_n^{1/4}/L_n\to0\), equivalently \(L_n=\omega(r_n^{1/4})\). A fixed grid leaves the unscaled interpolation error unchanged as \(n\) grows; after multiplication by \(r_n^{1/2}\), this fixed interpolation error can dominate the stochastic scale. The condition therefore rules out fixed grids for asymptotic validity while still allowing grids much smaller than rules that grow at order \(r_n^{1/2}\).

\begin{figure}[!htbp]
	\centering 
	\caption{Interpolation Error with $ F(x) = \cos(5x) + \sqrt{x}, \, x \in [0,5] $ }
	\begin{subfigure}{0.3\textwidth}
		\centering
		\includegraphics[width=1\textwidth]{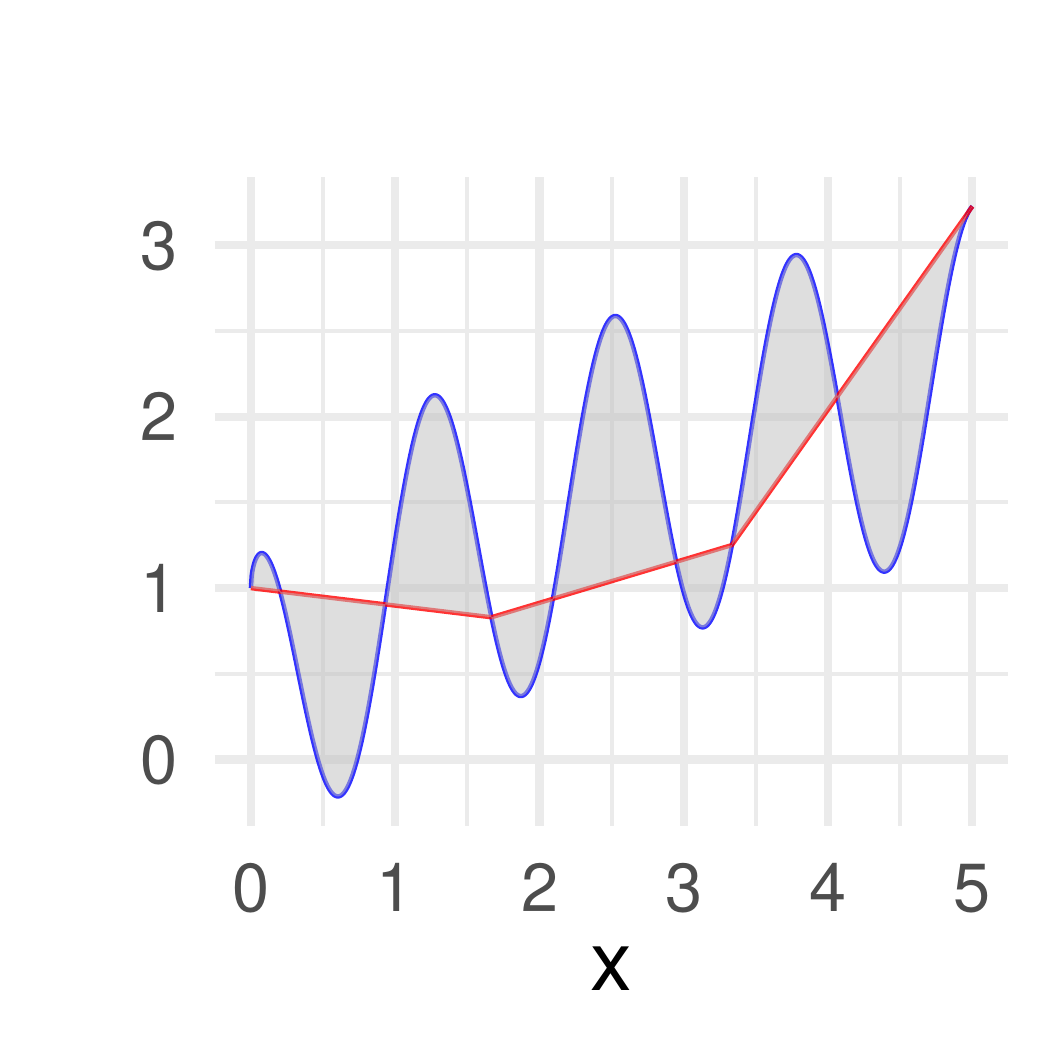}
			\caption{ $L= 4 $,\\ $a.x=3.441$, $m.x=1.542$ }
	\end{subfigure}
	\begin{subfigure}{0.3\textwidth}
		\centering
		\includegraphics[width=1\textwidth]{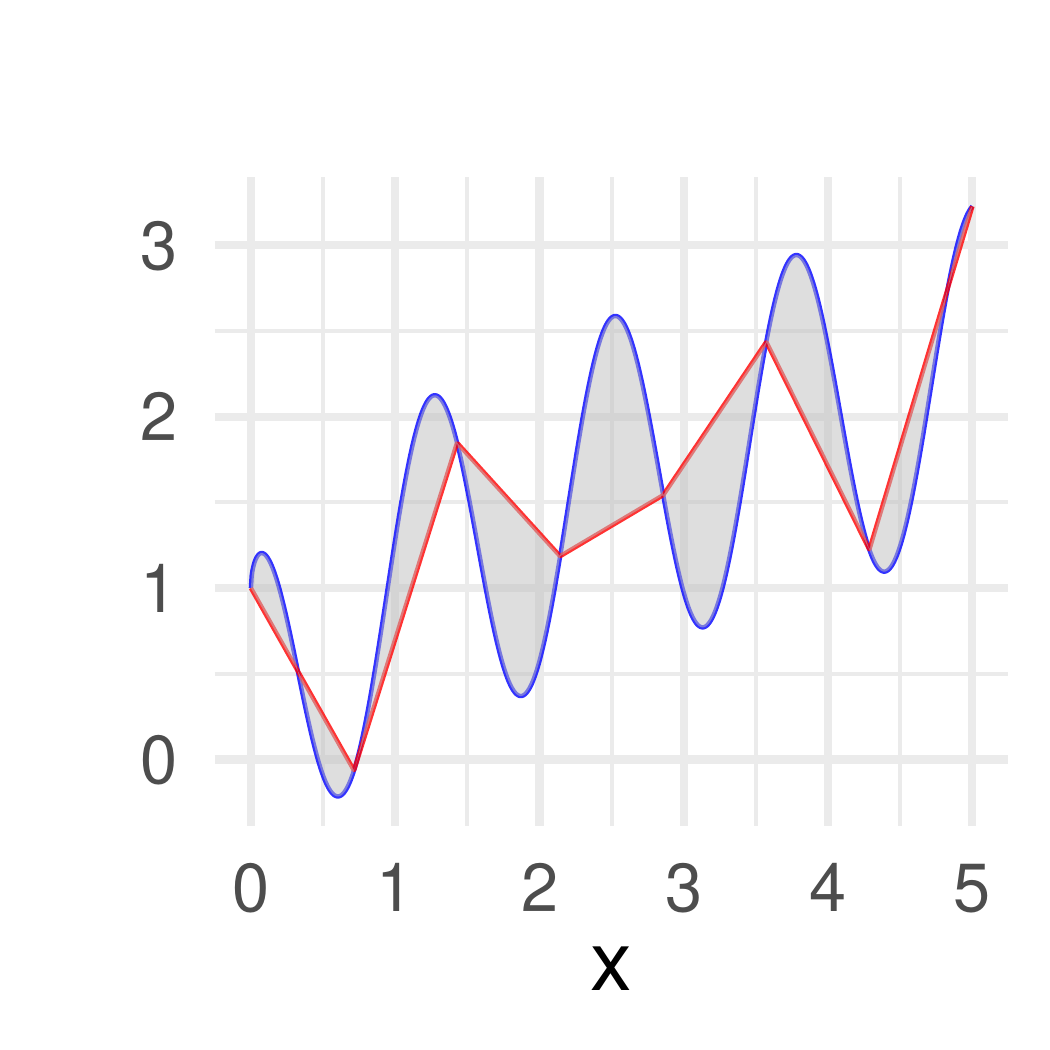}
			\caption{ $L= 8 $,\\ $a.x=2.608$, $m.x=1.215$  }
	\end{subfigure}
	\begin{subfigure}{0.3\textwidth}
		\centering
		\includegraphics[width=1\textwidth]{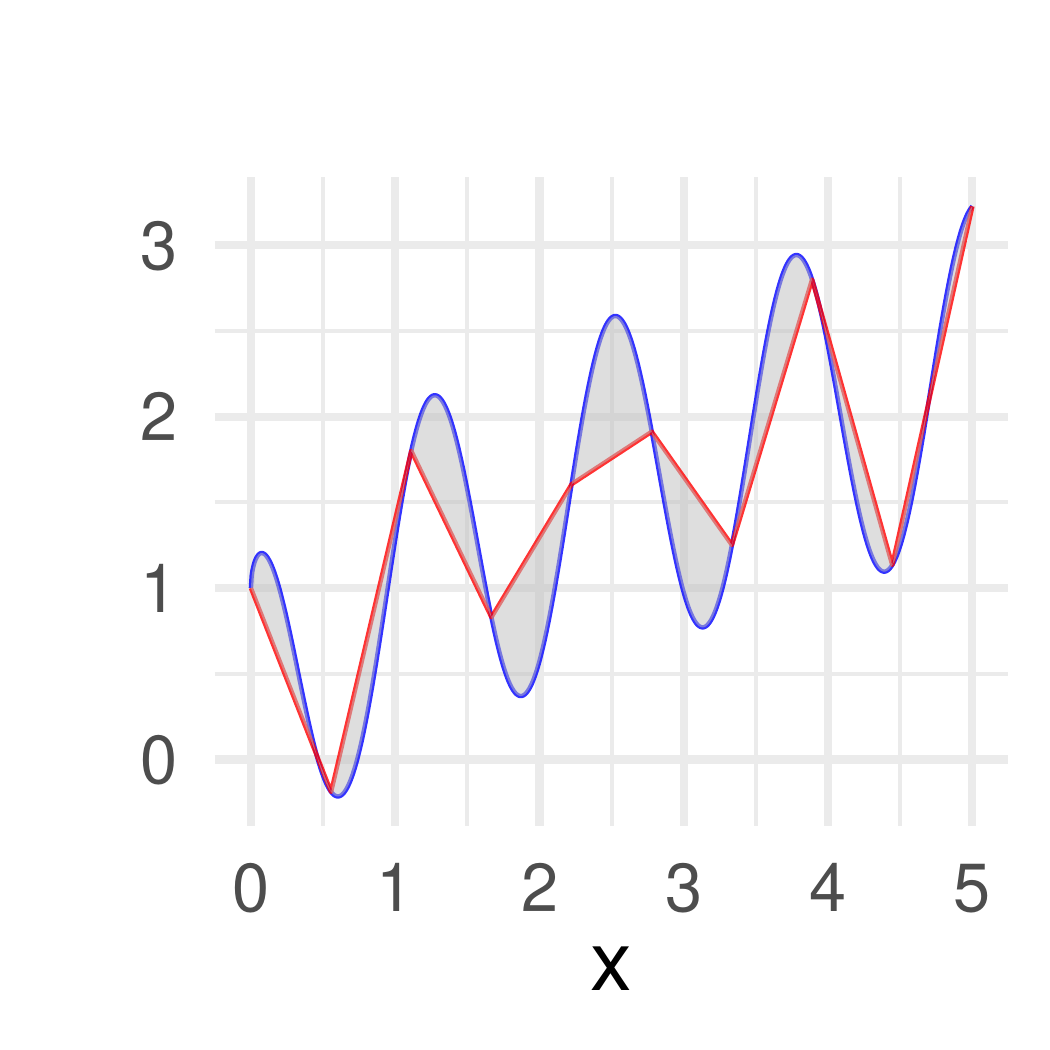}
			\caption{ $L= 10 $,\\ $a.x=1.720$, $m.x=0.820$ }
	\end{subfigure}
	\label{Fig:AE_illus_a}
	{\footnotesize 
		\begin{justify}
					The absolute interpolation error is shaded grey; \(a.x\) denotes its integrated area, and \(m.x\) denotes its maximum.
		\end{justify}
	}
\end{figure}

\Cref{Fig:AE_illus_a} illustrates the unscaled interpolation error from linearly interpolating the function \(F(x)=\cos(5x)+\sqrt{x}\). In this one-dimensional illustration, the shaded area corresponds to the integrated absolute interpolation error \( \displaystyle \int_\X |(F_L-F)(x)|\,dx \), while \(m.x\) denotes the maximum absolute interpolation error \( \displaystyle \sup_{x\in\X}|(F_L-F)(x)| \). Both measures shrink as \(L\) increases.

\begin{remark}\label{rem:lip_holder}
	For functions that are not twice continuously differentiable, such as Lipschitz or H\"older-continuous functions, the interpolation error is controlled by the modulus of continuity. For any continuous function $F$ on $\X$,
	\[
	\|F_L - F\|_{\infty}
	\;\le\;
	\sup_{\substack{\x,\y\in\X\\ \dmet(\x,\y) \le \delta_{\varepsilon}}}
	|F(\x)-F(\y)|,
	\qquad
	\delta_{\varepsilon}:=\max_{1\leq k\leq d}\varepsilon_k;
	\]
	see \citet[eqn. 3.2.8]{atkinson-han-2009}. If, in addition, \(F\) is H\"older-continuous with exponent \(\alpha\in(0,1]\) and constant \(\widetilde C\) (with Lipschitz continuity as the special case \(\alpha=1\)), then
	\[
	\sup_{\x \in \X} \big|R_n(\x) \big|
	\le
	r_n^{1/2}\widetilde{C}\,\delta_{\varepsilon}^{\alpha}
	\le
	\frac{r_n^{1/2}\,\widetilde{C}S_{\X}^{\alpha}}{(L-1)^\alpha}.
	\]
	Consequently, the rate condition $L_n = \omega(r_n^{1/(2\alpha)})$ ensures \( \displaystyle \sup_{\x\in\X}|R_n(\x)|=o(1) \) in this setting.
\end{remark}

The main results are stated under \Cref{cond:f_second_deriv_bound} because twice continuous differentiability gives the clean benchmark rate $L_n=\omega(r_n^{1/4})$. Twice-continuous-differentiability assumptions of this kind are common in continuous-index distributional and quantile-process settings; see, for example, \citet[Assumption 1 (2)]{hausman-liu-luo-palmer-2021errors} and \citet[Assumption 3(i)]{callaway-li-murt-tsyawo-2026distributional}. The display in \Cref{rem:lip_holder} indicates how the grid-rate condition changes under weaker H\"older smoothness.

\section{Inference}\label{sec:inference}
Having isolated the deterministic interpolation error \(R_n(\x)=r_n^{1/2}\{F_L(\x)-F(\x)\}\), this section studies its effect on inference for the interpolated estimator \(\widehat F_L\). The underlying centred empirical process is \(Z_n:=r_n^{1/2}(\widehat F-F)\). The continuous-index benchmark is infeasible because it requires evaluating the estimator, bootstrap process, and sup-\(t\) statistic over the continuum \(\X\), whereas computation can only be carried out on a finite tensor grid. This section first shows that interpolation preserves the weak limit of this process under the grid-growth condition, then records the computational algorithm, and finally proves that the resulting interpolated confidence bands have the same asymptotic coverage as the corresponding infeasible benchmark under the rate condition.

\subsection{Asymptotic Tightness and Weak Convergence}

The stochastic component of the analysis is formulated in a Donsker framework. This setting covers many continuous-index objects used in econometrics, including counterfactual distribution functions based on quantile and distribution regression \citep{chernozhukov-fernandez-melly-2013inference}, quantile and distributional treatment-effect processes in difference-in-differences settings \citep{callaway-li-2019,kim-wooldridge-2024-difference}, and multivariate distribution-regression processes \citep{meier-2025}. Let \(\mathbb G_n\psi_{\x}\) denote the centred empirical-process term in the asymptotic linear representation of \(Z_n(\x)\). In the usual i.i.d. \(\sqrt n\)-regular case, \( \displaystyle \mathbb G_n f:=n^{-1/2}\sum_{i=1}^n\{f(\mathcal W_i)-Pf\}\).
\begin{condition}[Donsker representation]\label{cond:donsker_repr}
	There exists a mean-zero class of influence functions
	\( \displaystyle 
	\Psi:=\{\psi_{\x}:\x\in\X\}
	\)
	such that \(\Psi\) is \(P\)-Donsker and is continuous in the index metric, in the sense that
	\( \displaystyle
	\lVert\psi_{\x}-\psi_{\y}\rVert_{P,2}
	\to0
	\qquad\text{whenever}\qquad
	\dmet(\x,\y)\to0
	\) where \(\lVert g\rVert_{P,2}:=(Pg^2)^{1/2}\).
	Furthermore,
	\[
	Z_n(\x)
	=
	\mathbb G_n\psi_{\x} + o_p(1)
	\qquad\text{in }\ell^\infty(\X).
	\]
	Let \(Z\) denote the tight Gaussian limit in \(\ell^\infty(\X)\) induced by the \(P\)-Donsker class \(\Psi\).
\end{condition}
\noindent \Cref{cond:donsker_repr} gives \(Z_n\rightsquigarrow Z\) in \(\ell^\infty(\X)\). In particular, it implies the finite-dimensional convergence and stochastic equicontinuity conditions, with respect to the index metric on \(\X\), that are usually used to establish weak convergence of an \(\R^d\)-indexed empirical process, e.g., \citet[Theorem 2.1]{Kosorok-2008-introduction}. 

The next result shows that the interpolated process inherits this weak limit once the deterministic interpolation error is asymptotically negligible.
\begin{theorem}\label{thm:weak_conv}
	If \Cref{cond:f_second_deriv_bound,cond:donsker_repr,cond:cond_compact_X} hold and \(L_n = \omega(r_n^{1/4})\), then \( \widehat{Z}_n^L \rightsquigarrow Z \text{ in } \ell^\infty(\X) \).
\end{theorem}\noindent \Cref{thm:weak_conv} shows that under the stated rate condition on the grid density over $\X$, the interpolated process converges weakly to the same limit as the \emph{infeasible} empirical process. The univariate result follows immediately by taking $d=1$.

\subsection{Computation algorithm}

The algorithm below abstracts the standard grid-based sup-$t$ construction for uniform confidence bands, as in \citet[Algorithm 3]{chernozhukov-fernandez-melly-2013inference} and \citet[Algorithm 1]{meier-2025}, and makes explicit the interpolation step that converts the grid-point bands into a continuous band over \(\X\). This records the standard operation that is otherwise implicit when estimates computed on a finite grid are plotted as continuous curves. Given grid-point bootstrap draws, a convenient scale estimator, following \citet{chernozhukov-fernandez-melly-2013inference} in the univariate case and \citet{meier-2025} in the multivariate case, is
\begin{align*}
    \widehat{\sigma}(\xbf) = \frac{\widehat Q_{\xbf}^*(0.75) - \widehat Q_{\xbf}^*(0.25)}{\Phi^{-1}(0.75)-\Phi^{-1}(0.25)}, \quad \xbf \in \X^L,
\end{align*}
where \(\widehat Q_{\xbf}^*(u)\) is the empirical \(u\)-quantile of \(\{Z_n^{*(b)}(\xbf):1\leq b\leq B\}\) and $\Phi^{-1}(\cdot)$ denotes the quantile function of the standard normal. Using the weights and vertices defined above, the interpolated estimator and scale process are then
\[
\widehat F_L(\x)
=
\sum_{\bm{\iota}\in\{0,1\}^d}
w_L(\x;\bm{\iota})\,\widehat F\big(\xbf_L(\x;\bm{\iota})\big)
\quad \text{and} \quad 
\widehat\sigma_L(\x)
=
\sum_{\bm{\iota}\in\{0,1\}^d}
w_L(\x;\bm{\iota})\,\widehat\sigma\big(\xbf_L(\x;\bm{\iota})\big),
\quad \x\in\X.
\]

\begin{algorithm}\caption{Sup-$t$ confidence bands for an interpolated functional $F_L$}\label{alg:supt_interp}
\quad
\begin{enumerate}
\item Fix a tensor-product grid $\X^L\subset\X$ and compute
\(\{\widehat F(\xbf):\xbf\in\X^L\}\).

\item Construct draws
\(\{Z_n^{*(b)}(\xbf):\xbf\in\X^L\}_{b=1}^B\)
from a conditionally weakly consistent estimator of the law of
\(r_n^{1/2}(\widehat F-F)\) on $\X^L$ and compute \(\widehat\sigma(\xbf)\) for each $\xbf\in\X^L$.

\item For each $b=1,\ldots,B$, compute
\[
T_n^{L,(b)}
:=
\max_{\xbf\in\X^L}
\frac{|Z_n^{*(b)}(\xbf)|}{\widehat\sigma(\xbf)}.
\]
Let \(t_{n,1-\alpha}^L:=\widehat Q_{T_n^L}(1-\alpha)\) be the empirical approximation to the conditional $(1-\alpha)$ quantile of the grid-bootstrap statistic, computed from \(\{T_n^{L,(b)}\}_{b=1}^B\).

\item For each $\x\in\X$, return the interpolated $\widehat F_L(\x)$ and confidence band
\[
\mathcal{C}_{n,1-\alpha}^L(\x)
=
\left[
\widehat F_L(\x)-\widehat\sigma_L(\x)\frac{t_{n,1-\alpha}^L}{r_n^{1/2}},
\quad \widehat F_L(\x)+\widehat\sigma_L(\x)\frac{t_{n,1-\alpha}^L}{r_n^{1/2}}
\right].
\]
\end{enumerate}
\end{algorithm}

Equivalently, by the linearity of the interpolation operator and because the same critical value \(t_{n,1-\alpha}^L\) is used at every grid point, the band in \Cref{alg:supt_interp} is obtained by interpolating the grid-point estimate and the lower and upper endpoints of the simultaneous grid-point bands.

Steps 2 and 3 of \Cref{alg:supt_interp} can be implemented using exchangeable or multiplier bootstrap draws. For exchangeable bootstrap schemes, \(Z_n^{*(b)}(\xbf)=r_n^{1/2}\{\widehat F^{*(b)}(\xbf)-\widehat F(\xbf)\}\). For multiplier schemes, \(Z_n^{*(b)}(\xbf)\) may be constructed directly from the corresponding weighted influence-function representation. In asymptotically linear settings, both forms can be represented through weights \(\xi_{1,n},\ldots,\xi_{n,n}\), independent of the data, satisfying the moment restrictions stated below; see \citet{praestgaard-wellner-1993} and \citet[Chap. 10]{Kosorok-2008-introduction}. The output of \Cref{alg:supt_interp} is therefore a function-valued estimate and a function-valued band, both continuous on $\X$ in-sample, obtained by interpolating the grid-point estimate and the scale estimate.

\subsection{Validity of confidence bands}
Since $ \widehat{F}_L $ is the reported continuous estimator used to construct the band, the inference procedure is completed with $\sup$-t confidence bands (CB) on $F$ using $ \widehat{F}_L $, which is continuous in $\x$ over $\X$. This subsection focuses on the multivariate case, which generalises the univariate case. Define the infeasible confidence band by
\[
\widetilde{\mathcal{C}}_{n,1-\alpha}(\x)
:=
\left[
\widehat{F}(\x)
\ - \
\widehat{\sigma}(\x)\,\frac{t_{n,1-\alpha}}{r_n^{1/2}}, \quad \widehat{F}(\x)
\ + \
\widehat{\sigma}(\x)\,\frac{t_{n,1-\alpha}}{r_n^{1/2}}
\right],
\quad \x \in \X
\]
where $ \widehat{\sigma}(\x) $ is the robust bootstrap standard-error function and \(t_{n,1-\alpha}:=\widehat Q_{T_{n,\infty}}(1-\alpha)\) for \( \displaystyle T_{n,\infty}:=\sup_{\x\in\X} \frac{|Z_n^*(\x)|}{\widehat{\sigma}(\x)} \). This is the continuous-index benchmark; interpolation of the feasible simultaneous confidence bands on the tensor grid $\X^L$ instead yields feasible continuous bands $\mathcal{C}_{n,1-\alpha}^L(\x), \ \x\in\X$.

In order to examine the asymptotic validity of the interpolated confidence bands $\mathcal{C}_{n,1-\alpha}^L$, the following imposes bootstrap approximation and robust studentisation conditions used to compute the sup-\(t\) critical value.
\begin{condition}[Bootstrap validity and robust studentisation]\label{cond:valid_CB_f}
Let \(\xi_{1,n},\ldots,\xi_{n,n}\) be bootstrap weights, independent of the data, and let \(\widehat{\Prob}\) denote probability under the conditional bootstrap law. Suppose the weights satisfy 
\[ \displaystyle
\frac{1}{n}\sum_{i=1}^n \xi_{i,n}
=
o_{\widehat{\Prob}}(1),
\quad
\frac{1}{n}\sum_{i=1}^n \xi_{i,n}^2
\xrightarrow{\widehat{\Prob}}
1,
\quad \text{and} \quad
\frac{1}{n}\sum_{i=1}^n |\xi_{i,n}|^{2+\delta}
=
\mathcal{O}_{\widehat{\Prob}}(1)
\]
in probability for some \(\delta>0\). Let
\( \displaystyle
\mathbb G_n^*\psi_{\x}
:=
\frac{1}{\sqrt n}\sum_{i=1}^n\xi_{i,n}
\psi_{\x}(\mathcal W_i),
\quad \x\in\X,
\)
where \(\mathcal W_i\) denotes the observed data. The bootstrap process admits the expansion
\( \displaystyle
Z_n^*(\x)
=
\mathbb G_n^*\psi_{\x}+o_{\widehat{\Prob}}(1)
\quad\text{in }\ell^\infty(\X),
\)
in probability, where \(\Psi=\{\psi_{\x}:\x\in\X\}\) is the \(P\)-Donsker class in \Cref{cond:donsker_repr}. The Gaussian limit \(Z\) in \Cref{cond:donsker_repr} has continuous sample paths on \(\X\), and its standard-error function, with \(\V[\cdot]\) denoting variance,
\(
\sigma(\x):=\{\V[Z(\x)]\}^{1/2}
\)
is continuous on \(\X\) and satisfies
\(\displaystyle
0<C^{-1}\leq \inf_{\x\in\X}\sigma(\x)
\leq \sup_{\x\in\X}\sigma(\x)\leq C<\infty .
\)
\end{condition}

Since \(\Psi\) is \(P\)-Donsker by \Cref{cond:donsker_repr}, the conditional weighted-bootstrap central limit theorem yields
\( \displaystyle 
\mathbb G_n^*\psi_{\cdot}
\rightsquigarrow_{\widehat{\Prob}}
Z
\quad\text{in }\ell^\infty(\X),
\)
see, e.g., \citet{praestgaard-wellner-1993} for exchangeably weighted bootstraps and \citet[Chap. 10]{Kosorok-2008-introduction} for the multiplier bootstrap. Together with the bootstrap expansion in \Cref{cond:valid_CB_f}, this gives
\(
Z_n^*
\rightsquigarrow_{\widehat{\Prob}}
Z
\quad\text{in }\ell^\infty(\X).
\)
By \citet[Supplementary Appendix, Lemma SA.1(a)--(b)]{chernozhukov-fernandez-melly-2013inference}, this conditional weak convergence implies
\[
\sup_{\x\in\X}\big|\widehat\sigma(\x)-\sigma(\x)\big|=o_p(1)
\quad \text{and} \quad 
W_n^*:=Z_n^*/\widehat\sigma
\rightsquigarrow_{\widehat{\Prob}}
W:=Z/\sigma
\quad\text{in }\ell^\infty(\X).
\]
Let \(t_{1-\alpha}:=Q_T(1-\alpha)\) be the population \((1-\alpha)\)-quantile of \( \displaystyle T:=\sup_{\x\in\X}|W(\x)|\). Then \(T\) has a continuous distribution at \(t_{1-\alpha}\), \(t_{n,1-\alpha}\to_p t_{1-\alpha}\), and
\( \displaystyle 
\lim_{n\rightarrow \infty}\Prob\Big[F(\x) \in \widetilde{\mathcal{C}}_{n,1-\alpha}(\x) \ \forall \ \x\in\X \Big] = (1-\alpha).
\)


The following result shows that the in-sample continuous uniform bands based on the interpolated process, which arise from the usual practice of displaying bands computed on a finite tensor grid $\X$ as continuous curves, have asymptotically valid coverage under the rate condition.
\begin{theorem}\label{thm:valid_CB}
 If \Cref{cond:cond_compact_X,cond:f_second_deriv_bound,cond:donsker_repr,cond:valid_CB_f} hold, then, under the rate condition $L_n=\omega(r_n^{1/4})$,
\[
    \lim_{n\rightarrow \infty}\Prob\Big[F(\x) \in \mathcal{C}_{n,1-\alpha}^L(\x) \ \forall \ \x\in\X \Big] = (1-\alpha).
\]
\end{theorem}
\noindent A main takeaway of the above result is that the interpolated continuous curves usually displayed as confidence bands are asymptotically valid under the rate condition $ L_n=\omega(r_n^{1/4}) $. This clarifies an often implicit step in the literature on uniform inference with estimators whose associated function classes are Donsker.

\subsection{Hadamard differentiable transformations}\label{rem:phi_extension}
Although the preceding results are stated for \(F\), they extend naturally to transformations \(\phi(F)\) whenever \(\phi\) is Hadamard differentiable at \(F\). The useful way to see this is through the decomposition in \Cref{eqn:rho_f_expand}, where \(Z_n^L\) is the interpolated stochastic component and \(R_n\) is the scaled deterministic interpolation error. Suppose \(\phi\) is Hadamard differentiable at \(F\), tangentially to \(\ell^\infty(\X)\), with continuous linear derivative \(\dot\phi_F\). Then the functional delta method, e.g., \citet[Theorem 20.8]{van-der-vaart-2000}, gives the scaled expansion
\[
r_n^{1/2}\{\phi(\widehat F_L)-\phi(F)\}
=
\dot\phi_F(\widehat{Z}_n^L)+o_p(1)
=
\dot\phi_F(Z_n^L)+\dot\phi_F(R_n)+o_p(1).
\]
By the first part of the proof of \Cref{thm:weak_conv}, \(Z_n^L\) has the same weak limit as the underlying empirical process under the rate condition \(L_n=\omega(r_n^{1/4})\). The second term is the transformed interpolation bias: under the same rate condition, \(\|R_n\|_\infty=o(1)\), so continuity of \(\dot\phi_F\) implies \(\dot\phi_F(R_n)=o(1)\). Thus the same grid-growth condition that makes interpolation negligible for \(F\) also makes it negligible after Hadamard differentiable transformations.

This extension covers, for example, conditional distribution functions generated by distribution-regression models, including multivariate-index versions such as
\[
F_{\bm X\mid W=w}(\x\mid w)=\Lambda\{w'\beta(\x)\},
\]
where \(\x\in\X\subset\R^d\) and \(\Lambda\) is a known link function; see, e.g., \citet{chernozhukov-fernandez-melly-2013inference} and \citet{meier-2025}. For fixed \(w\), this is a transformation of the coefficient process \(\beta(\cdot)\), namely
\[
[\phi_w(\beta)](\x)=\Lambda\{w'\beta(\x)\}, \quad \x\in\X.
\]
Under the usual smoothness of \(\Lambda\), \(\phi_w\) is Hadamard differentiable with derivative \([\dot\phi_{w,\beta}(h)](\x)=\lambda\{w'\beta(\x)\}w'h(\x)\), where \(\lambda=\Lambda'\). Hence the same interpolation argument applies to the resulting conditional CDF process indexed by \(\x\in\X\), with the univariate case nested by \(d=1\).

As a leading example, let $\phi$ denote the quantile map, and let $F_X$ be a strictly
increasing and continuously differentiable distribution function with derivative
$f_X$ such that $\displaystyle \inf_{\x\in\X} f_X(x)>0$. Fix $\tau\in(0,1)$ and write $x_\tau:=F_X^{-1}(\tau)$. Then under the rate
condition $L_n=\omega(r_n^{1/4})$,
\begin{align*}
r_n^{1/2}\big(\widehat{F}_{X,L}^{-1} - F_X^{-1}\big)(\tau)
=&
r_n^{1/2}\big(\phi(\widehat{F}_{X,L}) - \phi(F_X)\big)(\tau) \\
=&
\dot{\phi}_{F_X}\!\left(
r_n^{1/2}\big(\widehat{F}_{X,L} - F_X\big)
\right)(\tau)
+
o_p(1) \\
=&
-\frac{1}{f_X(x_\tau)}\,\widehat{Z}_n^L(x_\tau)
+
o_p(1),
\end{align*}
where $\widehat{Z}_n^L(x_\tau):=r_n^{1/2}\big(\widehat{F}_{X,L} - F_X\big)(x_\tau)$. It follows under the conditions of \Cref{thm:weak_conv} and the continuous mapping theorem that
\( \displaystyle 
r_n^{1/2}\big(\widehat{F}_{X,L}^{-1} - F_X^{-1}\big)(\tau)
\ \rightsquigarrow\
-\frac{1}{f_X(x_\tau)}\,Z(x_\tau).
\)

\section{Monte Carlo evidence}\label{subsec:mc_evidence}
This section provides Monte Carlo evidence on the finite-sample performance of the interpolation-based confidence band construction in \Cref{alg:supt_interp}.
The simulations are designed to illustrate the asymptotic mechanism and the computational savings of the theory-guided rule; they are not meant to suggest that fixed grids must fail sharply in every moderate-sample design.
The simulations use two-period difference-in-differences designs with univariate and bivariate continuously distributed outcomes. For each sample size \(n\in\{250,500,1000,1500\}\), treatment status is generated as \(D_i\sim\mathrm{Bernoulli}(1/2)\), and the time indicator is set as $t_i = \indicator{i \geq 0.5n} $, i.e., \(t_i\in\{0,1\}\) is assigned so that half of the observations lie in each period.
The common scalar index is
\[
\mu_i=\alpha+\beta D_i+\gamma t_i+\theta D_i t_i,
\qquad
(\alpha,\beta,\gamma,\theta)=(0.1,\, 0.2,\, -0.1,\, 0).
\]
In the univariate design, \(Y_i=\mu_i+\Eps_i\) with \(\Eps_i\sim \mathcal{N}(0,1)\).
In the bivariate design,
\( \displaystyle 
Y_i= (1, \, 1)'\mu_i + \Eps_i, \qquad
\Eps_i\sim \mathcal{N}_2(0,\Sigma), 
\qquad
\Sigma_{jk} = 0.5^{|j-k|}, \ j,k \in \{1,2\}.
\)
Thus, in both designs, the treatment effect is set to zero.

The procedure is applied to two target functionals.
The first is the Distribution Treatment Effect on the Treated (DTT),
\[
F^{\mathrm{DTT}}(\x)
:=
F_{Y(1)\mid D=1}(\x)-F_{Y(0)\mid D=1}(\x),
\]
which satisfies \(F^{\mathrm{DTT}}\equiv0\) under \(\theta=0\).
The second is the untreated counterfactual CDF for the treated group,
\[
F^{\mathrm{CF}}(\x):=F_{Y(0)\mid D=1}(\x).
\]
Under the identity-link distributional difference-in-differences specification, e.g., \citet{kim-wooldridge-2024-difference}, the grid-point estimators are
\[
\widehat F^{\mathrm{CF}}(\x)
=
\widehat F_{10}(\x)+\widehat F_{01}(\x)-\widehat F_{00}(\x)
\quad \text{and} \quad
\widehat F^{\mathrm{DTT}}(\x)
=
\widehat F_{11}(\x)-\widehat F^{\mathrm{CF}}(\x),
\]
where \(\widehat F_{dt}\) is the empirical CDF in group \(d \in \{0,1\} \) at period \(t \in \{0,1\} \). The population targets are used to evaluate coverage and error.

For each object \(m\in\{\mathrm{DTT},\mathrm{CF}\}\), estimation is carried out on an equally spaced grid \(\X_m^L\), and the grid-point estimates and bootstrap endpoints are interpolated over the evaluation region
\[
\X_m=\prod_{k=1}^d[\underline y_{m,k},\bar y_{m,k}].
\]
Let \(S_m=\bar y_{m,1}-\underline y_{m,1}\) denote the object-specific coordinate span of this region.
The grid specifications include a deliberately coarse fixed rule, \(L_{m,n}^{\mathrm{fix}}=5\), a theory-guided rule based on \Cref{lem:interp_error},
\[
L_{m,n}^{\mathrm{th}}
=
\left\lceil
1+
\kappa
S_m
\left\{
\frac{d\log\log(\mathrm e+n)}{8}
\right\}^{1/2}
n^{1/4}
\right\rceil,
\]
where \(\log\log(\mathrm e+n)\) supplies a slowly diverging factor beyond the boundary rate. The span \(S_m\) is known once the evaluation region \(\X_m\) is specified, while \(\kappa>0\) is a calibration constant absorbing curvature and other conservative constants in the bound. The reported simulations set \(\kappa=1\). The comparison also includes four power rules
\[
L_{m,n}^{\mathrm{pow}}(a,b)
=
\left\lceil aS_m n^b\right\rceil,
\qquad
(a,b)\in\{1,2\}\times\{0.30,0.35\}.
\]
All rules are truncated below at two nodes per coordinate to ensure that interpolation is well-defined.\footnote{This finite-sample convention is asymptotically immaterial.} The fixed rule is intentionally coarse from an asymptotic perspective because it holds the number of grid cells fixed as \(n\) increases; the comparison is therefore informative about how much of the finite-sample performance is driven by grid refinement rather than by sampling variation alone.

The evaluation regions are central regions of the relevant population outcome distributions used to construct each target functional.
In the univariate design, the DTT region is the central \(5\%\)--\(95\%\) interval of the equally weighted mixture of the four group-time outcome distributions, while the counterfactual region is the central \(5\%\)--\(95\%\) interval of the counterfactual distribution. In the bivariate design, the DTT region is the symmetric box \([ -c,c]^2\), where \(c\) is chosen so that the lower-orthant CDF of the equally weighted mixture satisfies \(F(c,c)-F(-c,-c)=0.9\); the counterfactual region is the diagonal box whose lower and upper corners have counterfactual lower-orthant probabilities \(0.05\) and \(0.95\), respectively.

Each design uses \(R=1000\) Monte Carlo replications and \(B=499\) ordinary non-parametric bootstrap draws.
For each object and grid rule, the bootstrap draws are used to form the grid-point sup-\(t\) critical value, and the resulting endpoints are interpolated to obtain the in-sample continuous band \(\mathcal{C}_{m,n,1-\alpha}^L\).
\Cref{Tab:mc_univariate,Tab:mc_bivariate} report two performance measures: the Monte Carlo coverage frequency of the continuous uniform-coverage event
\[
\indicatorBig{
	F^m(\x)\in\mathcal{C}_{m,n,1-\alpha}^L(\x)
	\ \text{for all }\x\in\X_m
}
=
\indicatorBigg{
	\sup_{\x\in\X_m}
	\frac{r_n^{1/2}\big|\widehat F_{m,L}(\x)-F^m(\x)\big|}
	{\widehat\sigma_{m,L}(\x)}
	\leq
	t_{m,n,1-\alpha}^L
},
\]
at nominal \(90\%\), \(95\%\), and \(99\%\) levels, and the normalised continuous error
\[
\mathcal L_{2,m}
=
\left(
\frac{1}{|\X_m|}
\int_{\X_m}
\big(\widehat F_{m,L}(\x)-F^m(\x)\big)^2\,d\x
\right)^{1/2},
\]
computed for the interpolated function over the full evaluation region.
In the baseline designs, the DTT continuous uniform-coverage event is especially direct because \(F^{\mathrm{DTT}}\equiv0\), while the counterfactual coverage and \(\mathcal L_2\) criteria use the corresponding population targets.

\begin{table}[!htbp]
\caption{\label{Tab:mc_univariate} Monte Carlo evidence, \(\R\)-indexed process}
\centering
\scriptsize
\setlength{\tabcolsep}{3pt}
\renewcommand{\arraystretch}{0.92}
\resizebox{\textwidth}{!}{%
\begin{tabular}{@{}clrrrrrrrrrr@{}}
\toprule
& & \multicolumn{5}{c}{DTT} & \multicolumn{5}{c}{CF} \\
\cmidrule(lr){3-7}\cmidrule(lr){8-12}
\(n\) & Rule & \(L\) & \(\mathcal L_2\) & \(90\%\) & \(95\%\) & \(99\%\) & \(L\) & \(\mathcal L_2\) & \(90\%\) & \(95\%\) & \(99\%\) \\
\midrule
\multirow{6}{*}{250} & Fixed & 5 & 0.083 & 0.876 & 0.941 & 0.991 & 5 & 0.072 & 0.880 & 0.944 & 0.989 \\
& Theory & 8 & 0.088 & 0.873 & 0.938 & 0.983 & 8 & 0.076 & 0.885 & 0.938 & 0.989 \\
& \((1,0.30)\) & 18 & 0.092 & 0.871 & 0.933 & 0.987 & 18 & 0.080 & 0.884 & 0.944 & 0.988 \\
& \((2,0.30)\) & 35 & 0.094 & 0.870 & 0.932 & 0.986 & 35 & 0.081 & 0.887 & 0.941 & 0.986 \\
& \((1,0.35)\) & 23 & 0.093 & 0.866 & 0.932 & 0.986 & 23 & 0.080 & 0.876 & 0.933 & 0.987 \\
& \((2,0.35)\) & 46 & 0.094 & 0.877 & 0.939 & 0.989 & 46 & 0.081 & 0.886 & 0.938 & 0.988 \\
\midrule
\multirow{6}{*}{500} & Fixed & 5 & 0.057 & 0.892 & 0.932 & 0.980 & 5 & 0.052 & 0.890 & 0.945 & 0.985 \\
& Theory & 9 & 0.062 & 0.895 & 0.934 & 0.984 & 9 & 0.054 & 0.898 & 0.949 & 0.990 \\
& \((1,0.30)\) & 22 & 0.065 & 0.882 & 0.939 & 0.987 & 22 & 0.056 & 0.894 & 0.946 & 0.986 \\
& \((2,0.30)\) & 43 & 0.066 & 0.888 & 0.938 & 0.988 & 43 & 0.057 & 0.899 & 0.943 & 0.987 \\
& \((1,0.35)\) & 30 & 0.065 & 0.894 & 0.940 & 0.986 & 29 & 0.056 & 0.895 & 0.952 & 0.990 \\
& \((2,0.35)\) & 59 & 0.066 & 0.890 & 0.940 & 0.988 & 58 & 0.057 & 0.894 & 0.948 & 0.985 \\
\midrule
\multirow{6}{*}{1000} & Fixed & 5 & 0.041 & 0.888 & 0.946 & 0.985 & 5 & 0.038 & 0.850 & 0.933 & 0.984 \\
& Theory & 11 & 0.045 & 0.887 & 0.949 & 0.985 & 11 & 0.039 & 0.901 & 0.942 & 0.989 \\
& \((1,0.30)\) & 27 & 0.047 & 0.879 & 0.944 & 0.986 & 27 & 0.040 & 0.895 & 0.953 & 0.990 \\
& \((2,0.30)\) & 53 & 0.047 & 0.887 & 0.937 & 0.988 & 53 & 0.040 & 0.896 & 0.945 & 0.988 \\
& \((1,0.35)\) & 38 & 0.047 & 0.882 & 0.940 & 0.987 & 37 & 0.040 & 0.897 & 0.951 & 0.990 \\
& \((2,0.35)\) & 75 & 0.047 & 0.882 & 0.940 & 0.988 & 74 & 0.041 & 0.893 & 0.947 & 0.987 \\
\midrule
\multirow{6}{*}{1500} & Fixed & 5 & 0.032 & 0.900 & 0.946 & 0.987 & 5 & 0.032 & 0.851 & 0.925 & 0.986 \\
& Theory & 12 & 0.036 & 0.892 & 0.956 & 0.988 & 12 & 0.031 & 0.916 & 0.956 & 0.991 \\
& \((1,0.30)\) & 30 & 0.037 & 0.903 & 0.954 & 0.993 & 30 & 0.032 & 0.898 & 0.953 & 0.991 \\
& \((2,0.30)\) & 60 & 0.038 & 0.887 & 0.954 & 0.991 & 60 & 0.032 & 0.908 & 0.955 & 0.995 \\
& \((1,0.35)\) & 43 & 0.037 & 0.899 & 0.950 & 0.989 & 43 & 0.032 & 0.910 & 0.956 & 0.992 \\
& \((2,0.35)\) & 86 & 0.038 & 0.892 & 0.946 & 0.992 & 86 & 0.033 & 0.897 & 0.958 & 0.994 \\
\bottomrule
\end{tabular}}
\begin{justify}
\footnotesize Notes: DTT denotes the distributional treatment effect on the treated and CF denotes the counterfactual distribution. The fixed rule sets \(L=5\) for every sample size; Theory uses the \(\kappa=1\) rule; \((a,b)\) indexes the power rule.
\end{justify}
\end{table}

\begin{table}[!htbp]
\caption{\label{Tab:mc_bivariate} Monte Carlo evidence, \(\R^2\)-indexed process}
\centering
\scriptsize
\setlength{\tabcolsep}{3pt}
\renewcommand{\arraystretch}{0.92}
\resizebox{\textwidth}{!}{%
\begin{tabular}{@{}clrrrrrrrrrr@{}}
\toprule
& & \multicolumn{5}{c}{DTT} & \multicolumn{5}{c}{CF} \\
\cmidrule(lr){3-7}\cmidrule(lr){8-12}
\(n\) & Rule & \(L\) & \(\mathcal L_2\) & \(90\%\) & \(95\%\) & \(99\%\) & \(L\) & \(\mathcal L_2\) & \(90\%\) & \(95\%\) & \(99\%\) \\
\midrule
\multirow{6}{*}{250} & Fixed & 5 & 0.074 & 0.926 & 0.971 & 0.994 & 5 & 0.079 & 0.888 & 0.942 & 0.985 \\
& Theory & 11 & 0.082 & 0.910 & 0.957 & 0.994 & 9 & 0.084 & 0.899 & 0.945 & 0.984 \\
& \((1,0.30)\) & 19 & 0.084 & 0.905 & 0.961 & 0.994 & 16 & 0.087 & 0.884 & 0.935 & 0.978 \\
& \((2,0.30)\) & 38 & 0.085 & 0.918 & 0.967 & 0.998 & 32 & 0.089 & 0.898 & 0.947 & 0.982 \\
& \((1,0.35)\) & 25 & 0.085 & 0.910 & 0.961 & 0.993 & 21 & 0.088 & 0.884 & 0.935 & 0.980 \\
& \((2,0.35)\) & 49 & 0.086 & 0.923 & 0.968 & 0.996 & 42 & 0.089 & 0.895 & 0.943 & 0.981 \\
\midrule
\multirow{6}{*}{500} & Fixed & 5 & 0.053 & 0.913 & 0.962 & 0.992 & 5 & 0.056 & 0.897 & 0.956 & 0.995 \\
& Theory & 13 & 0.058 & 0.927 & 0.968 & 0.993 & 11 & 0.060 & 0.896 & 0.949 & 0.990 \\
& \((1,0.30)\) & 23 & 0.060 & 0.913 & 0.961 & 0.993 & 20 & 0.062 & 0.885 & 0.945 & 0.989 \\
& \((2,0.30)\) & 46 & 0.061 & 0.925 & 0.969 & 0.994 & 39 & 0.062 & 0.908 & 0.958 & 0.994 \\
& \((1,0.35)\) & 32 & 0.060 & 0.925 & 0.975 & 0.992 & 27 & 0.062 & 0.894 & 0.948 & 0.992 \\
& \((2,0.35)\) & 63 & 0.061 & 0.932 & 0.974 & 0.994 & 54 & 0.062 & 0.906 & 0.958 & 0.993 \\
\midrule
\multirow{6}{*}{1000} & Fixed & 5 & 0.037 & 0.895 & 0.949 & 0.989 & 5 & 0.041 & 0.872 & 0.931 & 0.984 \\
& Theory & 15 & 0.042 & 0.926 & 0.966 & 0.986 & 13 & 0.043 & 0.905 & 0.957 & 0.992 \\
& \((1,0.30)\) & 29 & 0.043 & 0.894 & 0.950 & 0.988 & 24 & 0.044 & 0.886 & 0.937 & 0.983 \\
& \((2,0.30)\) & 57 & 0.043 & 0.918 & 0.959 & 0.990 & 48 & 0.044 & 0.897 & 0.956 & 0.994 \\
& \((1,0.35)\) & 40 & 0.043 & 0.916 & 0.955 & 0.990 & 34 & 0.044 & 0.906 & 0.953 & 0.991 \\
& \((2,0.35)\) & 80 & 0.044 & 0.925 & 0.960 & 0.990 & 68 & 0.045 & 0.908 & 0.958 & 0.993 \\
\midrule
\multirow{6}{*}{1500} & Fixed & 5 & 0.031 & 0.910 & 0.960 & 0.992 & 5 & 0.035 & 0.845 & 0.913 & 0.978 \\
& Theory & 17 & 0.035 & 0.897 & 0.948 & 0.988 & 15 & 0.036 & 0.902 & 0.946 & 0.987 \\
& \((1,0.30)\) & 32 & 0.035 & 0.921 & 0.959 & 0.993 & 28 & 0.037 & 0.884 & 0.934 & 0.982 \\
& \((2,0.30)\) & 64 & 0.036 & 0.919 & 0.955 & 0.994 & 55 & 0.037 & 0.908 & 0.946 & 0.989 \\
& \((1,0.35)\) & 46 & 0.035 & 0.914 & 0.954 & 0.993 & 40 & 0.037 & 0.904 & 0.940 & 0.988 \\
& \((2,0.35)\) & 92 & 0.036 & 0.918 & 0.957 & 0.992 & 79 & 0.037 & 0.913 & 0.948 & 0.988 \\
\bottomrule
\end{tabular}}
\begin{justify}
\footnotesize Notes: DTT denotes the distributional treatment effect on the treated and CF denotes the counterfactual distribution. The fixed rule sets \(L=5\) along each coordinate for every sample size; Theory uses the \(\kappa=1\) rule; \((a,b)\) indexes the power rule.
\end{justify}
\end{table}

The simulations are broadly reassuring but also illustrate the intended contrast between fixed and increasing grids. Across both the univariate and bivariate designs, the increasing-grid rules deliver coverage frequencies that are generally close to the nominal levels. In the univariate design, for example, the theory-guided rule attains \(95\%\) coverage between \(0.934\) and \(0.956\) for DTT and between \(0.938\) and \(0.956\) for the counterfactual CDF across the four sample sizes. In the bivariate design, the corresponding \(95\%\) coverage ranges are \(0.948\)--\(0.968\) for DTT and \(0.945\)--\(0.957\) for the counterfactual CDF.

The deliberately coarse fixed rule, \(L=5\), performs reasonably at the smallest sample sizes, but its coverage is weaker in several cells, especially for the counterfactual distribution. This rule should not be read as a claim that practitioners typically use only five nodes. Rather, it is a transparent finite-sample diagnostic for the asymptotic mechanism: for any fixed \(L\), however large, the interpolation bias remains of order \(L^{-2}\), whereas sampling uncertainty shrinks with \(n\). Hence, sufficiently large samples eventually expose the fixed-grid interpolation error; the role of \(L=5\) is simply to make this mechanism visible at the sample sizes considered. For instance, in the univariate counterfactual design, \(95\%\) coverage under the fixed rule falls from \(0.944\) at \(n=250\) to \(0.925\) at \(n=1500\); in the bivariate counterfactual design it declines from \(0.956\) at \(n=500\) to \(0.913\) at \(n=1500\). This pattern is consistent with the theory: a fixed grid need not fail sharply in moderate samples when the target functions are smooth, but it does not remove the interpolation bias asymptotically. \Cref{app:mc_stress} reports an additional smooth stress design in which this fixed-grid deterioration is innocuous at small sample sizes but more visible at larger sample sizes.
The relatively strong performance of the fixed rule for baseline DTT should be interpreted in light of the design: since \(F^{\mathrm{DTT}}\equiv0\), the population target is flat, so there is no curvature in the DTT target itself for a coarse grid to miss. The counterfactual CDF, and the non-zero DTT target in the stress design reported in \Cref{app:mc_stress}, are therefore more informative about interpolation error for curved targets.

The theory-guided rule is competitive in these designs. It typically delivers coverage and continuous \(\mathcal L_2\) error similar to the power rules while using fewer grid points, particularly in the bivariate design where the computational cost grows with \(L^d\). For example, at \(n=1500\) in the bivariate DTT design, the theory-guided rule uses \(L=17\) nodes per coordinate and attains \(95\%\) coverage of \(0.948\), while the largest power rule uses \(L=92\) and attains \(0.957\). The evidence therefore supports the practical message of the paper: grid refinement matters for uniform inference, but a rate-guided rule can avoid unnecessarily dense grids while preserving reliable finite-sample performance.

\section{Conclusion}\label{sect:concl}
This paper provides a theoretical basis for choosing the grid density over the index set $\X$ when conducting uniform inference on a continuous target functional $F$. The starting point is the practical observation that feasible simultaneous confidence bands are computed on a finite tensor grid but are typically reported as continuous curves by linearly interpolating between grid points. Treating this interpolation step as an integral part of the inferential procedure, rather than as a purely graphical convention, makes the interpolation error explicit and thereby provides the basis for choosing \(L_n\) so that uniform inference remains asymptotically valid.

Under twice continuous differentiability and weak convergence of the underlying empirical process, the paper shows that the interpolation error is asymptotically negligible whenever $L_n = \omega(r_n^{1/4})$. This rate condition is sufficient for the interpolated estimator to inherit the weak limit of the grid-point process and for the resulting interpolated confidence bands to attain their nominal uniform coverage. For estimators with the usual $r_n=n$ rate, the rule becomes $L_n=\omega(n^{1/4})$, giving a simple and theoretically grounded guide for balancing computational cost against asymptotic validity.

\vspace{0.75cm}

\noindent \texttt{Declaration of AI use:}
During the preparation of this work, the author used OpenAI's ChatGPT and Codex for language editing, manuscript consistency checks, LaTeX troubleshooting, and assistance with simulation-code organisation. The author reviewed and edited the output as needed and takes full responsibility for the content of the published article.
 
\printbibliography

\newpage
\hypersetup{pageanchor=false}
\setcounter{page}{1}
\appendix
\hypersetup{pageanchor=true}
\renewcommand{\thetable}{S.\arabic{table}}
\renewcommand{\thefigure}{S.\arabic{figure}}
\renewcommand{\thesection}{S.\arabic{section}}
\setcounter{equation}{0}
\renewcommand{\theequation}{S.\arabic{equation}}

\begin{center}
{\Large\bfseries Appendix}
\end{center}

\section{Proofs of Results}

\subsection[Proof of Lemma \ref{lem:interp_error}]{Proof of \Cref{lem:interp_error}}

By the multilinear interpolation error bound of \citet[Theorem 2.2]{weiser-1988-note}, applied with \(N=d\) and \( \displaystyle h=\max_{1\leq k\leq d}\varepsilon_k\), and noting that the univariate case reduces to the standard linear-interpolation remainder obtained from Rolle's theorem, e.g., \citet[Theorem 2.1]{epperson-2013-introduction}, it follows under \Cref{cond:cond_compact_X,cond:f_second_deriv_bound} that
\begin{align*}
    \sup_{\x \in \X}\big|F(\x)-F_L(\x)\big|
    &\leq \frac{d}{8} \Big(\max_{1\leq k \leq d} \varepsilon_k^2 \Big)
    \sup_{\bm{x} \in \X} \Big(\max_{1\leq k \leq d} \Big| \frac{\partial^2F(\bm{x})}{\partial x_k^2} \Big|\Big) \\
    &= \frac{d}{8}\max_{1\leq k \leq d} \Big(\frac{\bar{\xrm}_k - \underline{\xrm}_k}{L-1}\Big)^2 \sup_{\bm{x} \in \X} \Big(\max_{1\leq k \leq d} \Big| \frac{\partial^2F(\bm{x})}{\partial x_k^2} \Big|\Big) \\
    & \leq \frac{dM_F S_{\X}^2}{8(L-1)^2}.
\end{align*}
Thus,
\[
\sup_{\x \in \X } |R_n(\x)| = r_n^{1/2}\sup_{\x \in \X } |\big(F_L-F\big)(\x)| \leq \frac{dM_F S_{\X}^2}{8}\left(\frac{r_n^{1/4}}{L-1}\right)^2.
\]

Setting $L_n = \omega(r_n^{1/4}) $, \(L_n\to\infty\) and \((L_n-1)/L_n\to1\), so
\[
	\frac{dM_F S_{\X}^2}{8}
	\Big(\frac{r_n^{1/4}}{L_n-1}\Big)^2
	\rightarrow 0
\]
as $n\rightarrow \infty$.

\qed

\subsection[Proof of Theorem \ref{thm:weak_conv}]{Proof of \Cref{thm:weak_conv}}

\textbf{Part (a):} First, compare the interpolated empirical process \(Z_n^L\) with the infeasible process \(Z_n\). Let \(\delta_{\varepsilon_n}:=\max_{1\leq k\leq d}\varepsilon_{n,k}\) denote the maximum mesh width of the grid, where \(\varepsilon_{n,k}\) is the mesh \(\varepsilon_k\) evaluated at \(L=L_n\).
For every \(\x\in\X\), the interpolation weights are non-negative and sum to one, and every vertex \(\xbf_L(\x;\bm{\iota})\) lies in the same grid cell as \(\x\). Hence
\begin{align*}
\big|Z_n^L(\x)-Z_n(\x)\big|
&=
\left|
\sum_{\bm{\iota}\in\{0,1\}^d}
w_L(\x;\bm{\iota})
\left\{Z_n\big(\xbf_L(\x;\bm{\iota})\big)-Z_n(\x)\right\}
\right| \\
&\leq
\sup_{\dmet(\x,\y)\leq \delta_{\varepsilon_n}}
\big|Z_n(\y)-Z_n(\x)\big|.
\end{align*}
Taking the supremum over \(\x\in\X\) yields
\[
\lVert Z_n^L-Z_n\rVert_\infty
\leq
\sup_{\dmet(\x,\y)\leq \delta_{\varepsilon_n}}
\big|Z_n(\y)-Z_n(\x)\big|.
\]
By \Cref{cond:cond_compact_X}, \(L_n\to\infty\) implies \(\delta_{\varepsilon_n}\to0\). The \(P\)-Donsker property of \(\Psi\), together with the continuity of \(\x\mapsto\psi_{\x}\) in the \(L_2(P)\) metric imposed in \Cref{cond:donsker_repr}, implies stochastic equicontinuity of \(\mathbb G_n\psi_{\cdot}\) with respect to \(\dmet\) on \(\X\). Since \(Z_n=\mathbb G_n\psi_{\cdot}+o_p(1)\) in \(\ell^\infty(\X)\), the same stochastic equicontinuity holds for \(Z_n\). Therefore,
\(
\lVert Z_n^L-Z_n\rVert_\infty=o_p(1)
\) under the rate condition. Combining this with \(Z_n\rightsquigarrow Z\) from \Cref{cond:donsker_repr} gives
\[
Z_n^L\rightsquigarrow Z
\qquad\text{in }\ell^\infty(\X)
\]
by Slutsky's theorem under the rate condition.

\textbf{Part (b):} It remains to transfer the weak convergence from \(Z_n^L\) to \(\widehat{Z}_n^L\). By the decomposition in \Cref{eqn:rho_f_expand} and \Cref{lem:interp_error},
\[
\lVert \widehat{Z}_n^L-Z_n^L\rVert_\infty
=
\sup_{\x\in \X}|R_n(\x)|
=o(1)
\]
under the rate condition \(L_n=\omega(r_n^{1/4})\). Since \(Z_n^L\rightsquigarrow Z\) from part (a) above in \(\ell^\infty(\X)\), another application of Slutsky's theorem in \(\ell^\infty(\X)\) yields \(\widehat{Z}_n^L\rightsquigarrow Z\) under the rate condition.

\qed

\subsection[Proof of Theorem \ref{thm:valid_CB}]{Proof of \Cref{thm:valid_CB}}
\begin{align*}
    \Big\{ F(\x) \in \widetilde{\mathcal{C}}_{n,1-\alpha}(\x) \ \forall \ \x\in\X \Big\} &= \Big\{ \widehat{F}(\x) - \widehat{\sigma}(\x)\frac{t_{n,1-\alpha}}{r_n^{1/2}} \leq F(\x) \leq  \widehat{F}(\x) + \widehat{\sigma}(\x)\frac{t_{n,1-\alpha}}{r_n^{1/2}} \quad \forall \x\in\X \Big\}\\
    &= \Big\{ \frac{|r_n^{1/2}(\widehat{F}-F)(\x)|}{\widehat{\sigma}(\x)} \leq t_{n,1-\alpha}  \quad \forall \x\in\X \Big\}\\
    &= \Big\{ \sup_{\x\in\X} \frac{|r_n^{1/2}(\widehat{F}-F)(\x)|}{\widehat{\sigma}(\x)} \leq t_{n,1-\alpha} \Big\} =: \Big\{ \sup_{\x\in\X} \frac{|Z_n(\x)|}{\widehat{\sigma}(\x)} \leq t_{n,1-\alpha} \Big\}
\end{align*}
Similarly,
\begin{align*}
    \Big\{ F(\x) \in \mathcal{C}_{n,1-\alpha}^L(\x) \ \forall \ \x\in\X \Big\} &= \Big\{ \widehat{F}_L(\x) - \widehat{\sigma}_L(\x)\frac{t_{n,1-\alpha}^L}{r_n^{1/2}} \leq F(\x) \leq \widehat{F}_L(\x) + \widehat{\sigma}_L(\x)\frac{t_{n,1-\alpha}^L}{r_n^{1/2}} \quad \forall \x\in\X \Big\}\\
    &= \Big\{ \frac{|r_n^{1/2}(\widehat{F}_L -F)(\x)|}{\widehat{\sigma}_L(\x)} \leq t_{n,1-\alpha}^L  \quad \forall \x\in\X \Big\}\\
    &= \Big\{ \sup_{\x\in\X} \frac{|r_n^{1/2}(\widehat{F}_L-F)(\x)|}{\widehat{\sigma}_L(\x)} \leq t_{n,1-\alpha}^L \Big\} =: \Big\{ \sup_{\x\in\X} \frac{|\widehat{Z}_n^L(\x)|}{\widehat{\sigma}_L(\x)} \leq t_{n,1-\alpha}^L \Big\}.
\end{align*}

\noindent The proof proceeds by showing that both events above are asymptotically equivalent under the rate condition $L_n=\omega(r_n^{1/4})$, and then invoking the coverage consequence of \Cref{cond:valid_CB_f} to conclude. 

For arbitrary real numbers \(z^L,z,t^L,t\), the following elementary bound separates the effect of changing the statistic from the effect of changing the cutoff:
\begin{align*}
    \big| \indicator{z^L \leq t^L} &- \indicator{z \leq t} \big| \leq \big| \indicator{z^L \leq t^L} - \indicator{z^L \leq t} \big| + \big| \indicator{z^L \leq t} - \indicator{z \leq t} \big| \\
    &= \Big( \indicator{ t < z^L \leq t^L } + \indicator{ t^L < z^L \leq t } \Big) + \Big( \indicator{ z^L \leq t < z } + \indicator{ z \leq t < z^L } \Big) \\
    &= \indicatorbig{ (t^L\wedge t) < z^L \leq (t^L\vee t) } + \indicatorbig{ (z^L\wedge z) \leq t < (z^L\vee z) } \\
    &= \indicatorbig{ ((z^L-z)\wedge 0) \leq t-z < ((z^L-z)\vee 0) } + \indicatorbig{ ((t^L-t)\wedge 0) < z^L-t \leq ((t^L-t)\vee 0) }.
\end{align*}
The first line follows by adding and subtracting $\indicator{z^L \leq t}$ and applying the triangle inequality. The second line holds because each absolute-value term in the first line equals 1 exactly when the two indicators being compared disagree (and equals zero otherwise). For the cutoff-change term, for example,
\[
\big| \indicator{z^L \leq t^L} - \indicator{z^L \leq t} \big| = 1
\]
if and only if
\[
\big\{z^L \leq t^L \text{ and } z^L > t\big\}
\quad \text{or} \quad
\big\{z^L > t^L \text{ and } z^L \leq t\big\}.
\]
The third line rewrites the two cases using \(\wedge\) and \(\vee\). The fourth line expresses the same events after centring by \(z\) and \(t\), respectively, and then reorders the two summands.

Applying this elementary bound with
\[
z^L=\widehat T_n^L,\quad
z=T_n,\quad
t^L=t_{n,1-\alpha}^L,\ \text{and} \
t=t_{n,1-\alpha},
\]
yields
\begin{align}
    \Big| \indicatorBig{ \sup_{\x\in\X} \frac{|\widehat{Z}_n^L(\x)|}{\widehat{\sigma}_L(\x)} \leq & t_{n,1-\alpha}^L } - \indicatorBig{ \sup_{\x\in\X} \frac{|Z_n(\x)|}{\widehat{\sigma}(\x)} \leq t_{n,1-\alpha}} \Big|=: \Big| \indicatorbig{\widehat{T}_n^L \leq t_{n,1-\alpha}^L} - \indicatorbig{ T_n \leq t_{n,1-\alpha} } \Big| \nonumber \\
    \leq & \indicatorBig{ (\widehat{T}_n^L-T_n) \wedge 0 \, \leq \, (t_{n,1-\alpha} - T_n) \, < \, (\widehat{T}_n^L-T_n) \vee 0 } \nonumber \\
    &+ \indicatorBig{ (t_{n,1-\alpha}^L - t_{n,1-\alpha})\wedge 0 \, < \, (\widehat{T}_n^L - t_{n,1-\alpha}) \, \leq \, (t_{n,1-\alpha}^L - t_{n,1-\alpha}) \vee 0 }. \label{eqn:decomp_indicat}
\end{align}
Parts (a) and (b) of the proof below are dedicated to studying the summands of \eqref{eqn:decomp_indicat}.

\textbf{Part (a):} Consider the first summand of \eqref{eqn:decomp_indicat}, namely \( \displaystyle \indicatorBig{ (\widehat{T}_n^L-T_n) \wedge 0 \, \leq \, (t_{n,1-\alpha} - T_n) \, < \, (\widehat{T}_n^L-T_n) \vee 0 } \). Since the $\sup$ operator is $1$-Lipschitz in the uniform norm, and since \(\big|\,|a|-|b|\,\big|\leq |a-b|\) by the reverse triangle inequality, one has
\begin{align*}
    \big|\widehat{T}_n^L - T_n \big| = \bigg| \sup_{\x\in\X} \frac{|\widehat{Z}_n^L(\x)|}{\widehat{\sigma}_L(\x)} - \sup_{\x\in\X} \frac{|Z_n(\x)|}{\widehat{\sigma}(\x)} \bigg| &\leq \sup_{\x\in\X} \bigg| \frac{|\widehat{Z}_n^L(\x)|}{\widehat{\sigma}_L(\x)} - \frac{|Z_n(\x)|}{\widehat{\sigma}(\x)} \bigg| \\
    &\leq \sup_{\x\in\X} \bigg| \frac{\widehat{Z}_n^L(\x)}{\widehat{\sigma}_L(\x)} - \frac{Z_n(\x)}{\widehat{\sigma}(\x)} \bigg|.
\end{align*}
Next, consider the pointwise decomposition:
\begin{align*}
    \frac{\widehat{Z}_n^L(\x)}{\widehat{\sigma}_L(\x)} - \frac{Z_n(\x)}{\widehat{\sigma}(\x)} &= \frac{\widehat{Z}_n^L(\x) - Z_n(\x)}{\widehat{\sigma}(\x)} -  \big( \widehat{\sigma}_L(\x) - \widehat{\sigma}(\x) \big)\bigg( \frac{\widehat{Z}_n^L(\x)}{\widehat{\sigma}_L(\x)\widehat{\sigma}(\x)} \bigg).
\end{align*}
Applying the triangle inequality to this pointwise decomposition and then taking suprema gives
\begin{align}\label{eqn:T_n_conv}
    \big|\widehat{T}_n^L - T_n \big| \leq \sup_{\x\in\X} \frac{|\widehat{Z}_n^L(\x) - Z_n(\x)|}{\widehat{\sigma}(\x)} + \sup_{\x\in\X} \left\{\frac{|\widehat{Z}_n^L(\x)|}{\widehat{\sigma}_L(\x)\widehat{\sigma}(\x)} \big| \widehat{\sigma}_L(\x)-\widehat{\sigma}(\x) \big|\right\}.
\end{align}
By the decomposition in \Cref{eqn:rho_f_expand}, \( \widehat{Z}_n^L(\x) = Z_n^L(\x) + R_n(\x) \). Since the interpolation weights sum to one, it follows that for any $\x\in \X$,
\begin{align*}
 \widehat{Z}_n^L(\x) - Z_n(\x) &=  Z_n^L(\x) - Z_n(\x)  + R_n(\x)\\
&= \sum_{\bm{\iota}\in\{0,1\}^d}
w_L(\x;\bm{\iota})\,\Big\{Z_n\big(\xbf_L(\x;\bm{\iota})\big) - Z_n(\x) \Big\} + R_n(\x).
\end{align*}
Then, by the triangle inequality, using the fact that the vertices of a cell $\mathfrak{C}_{\bm{\ell}}$ containing a point $\x\in\X$ also belong to the cell, and that interpolation weights sum to one,
\begin{align*}
    \sup_{\x\in\X} \big|\widehat{Z}_n^L(\x) - Z_n(\x)\big| \leq \sup_{\lVert\x - \x'\rVert_{\infty} \leq \delta_{\varepsilon_n}} \big| Z_n(\x) - Z_n(\x') \big| + \sup_{\x\in\X} \big| R_n(\x) \big| = o_p(1)
\end{align*}
thanks to the stochastic equicontinuity implied by \Cref{cond:donsker_repr} and the rate condition $L_n=\omega(r_n^{1/4})$ under which $ \delta_{\varepsilon_n} = o(1) $ and $ \displaystyle \sup_{\x\in\X} \big| R_n(\x) \big| = o(1) $. Combining this with the robust scale consistency and non-degeneracy consequences of \Cref{cond:valid_CB_f}, one has
\begin{align*}
    \sup_{\x\in\X} \frac{|\widehat{Z}_n^L(\x) - Z_n(\x)|}{\widehat{\sigma}(\x)} = o_p(1)
\end{align*}
under the rate condition $L_n=\omega(r_n^{1/4})$.

Next, since the interpolation weights sum to one,
\begin{align*}
    \widehat{\sigma}_L(\x)-\widehat{\sigma}(\x)
    =& \sum_{\bm{\iota}\in\{0,1\}^d}
    w_L(\x;\bm{\iota})\,
    \Big\{\widehat{\sigma}\big(\xbf_L(\x;\bm{\iota})\big)-\widehat{\sigma}(\x)\Big\} \\
    =& \sum_{\bm{\iota}\in\{0,1\}^d}
    w_L(\x;\bm{\iota})\,
    \Big\{\sigma\big(\xbf_L(\x;\bm{\iota})\big)-\sigma(\x)\Big\} \\
    &+ \sum_{\bm{\iota}\in\{0,1\}^d}
    w_L(\x;\bm{\iota})\,
    \Big[
    \big\{\widehat{\sigma}\big(\xbf_L(\x;\bm{\iota})\big)-\sigma\big(\xbf_L(\x;\bm{\iota})\big)\big\}
    -
    \big\{\widehat{\sigma}(\x)-\sigma(\x)\big\}
    \Big].
\end{align*}
It follows by the triangle inequality that
\begin{align*}
    \big|\widehat{\sigma}_L(\x)-\widehat{\sigma}(\x)\big|
    \leq&
    \sum_{\bm{\iota}\in\{0,1\}^d}
    w_L(\x;\bm{\iota})\,
    \big|\sigma\big(\xbf_L(\x;\bm{\iota})\big)-\sigma(\x)\big| \\
    &+
    2\sup_{\x\in\X}\big|\widehat{\sigma}(\x)-\sigma(\x)\big|.
\end{align*}
Because $\sigma$ is continuous on the compact set $\X$, it is uniformly continuous. Therefore, under the shrinking-cell condition $\delta_{\varepsilon_n}=o(1)$ and the robust scale consistency consequence of \Cref{cond:valid_CB_f},
\[
\sup_{\x\in\X}\big|\widehat{\sigma}_L(\x)-\widehat{\sigma}(\x)\big|=o_p(1).
\]
Moreover, the robust scale consistency and non-degeneracy consequences of \Cref{cond:valid_CB_f} imply that $\widehat{\sigma}(\x)$ and $\widehat{\sigma}_L(\x)$ are bounded away from zero uniformly in $\x$ with probability approaching one. Since \(\widehat{Z}_n^L\rightsquigarrow Z\) in \(\ell^\infty(\X)\) by \Cref{thm:weak_conv}, \(\lVert \widehat{Z}_n^L\rVert_\infty=\Op(1)\). Hence,
\[
\sup_{\x\in\X}
\frac{|\widehat{Z}_n^L(\x)|}{\widehat{\sigma}_L(\x)\widehat{\sigma}(\x)}
\big|\widehat{\sigma}_L(\x)-\widehat{\sigma}(\x)\big|
=o_p(1).
\]
Thus, by \eqref{eqn:T_n_conv}, $ \widehat{T}_n^L - T_n = o_p(1) $ under the rate condition $L_n=\omega(r_n^{1/4})$. The indicator in Part (a) records the event that the benchmark cut-off falls between \(T_n\) and \(\widehat T_n^L\). Equivalently, it is bounded above by
\[
\indicatorBig{
T_n\wedge \widehat T_n^L
\leq
t_{n,1-\alpha}
<
T_n\vee \widehat T_n^L
}
\leq
\indicatorBig{|T_n-t_{n,1-\alpha}|\leq |\widehat T_n^L-T_n|}.
\]
Since \( |\widehat T_n^L-T_n|=o_p(1) \), there exists a deterministic sequence \(\eta_n\downarrow 0\) such that
\(\Prob(|\widehat T_n^L-T_n|>\eta_n)\to 0\). By \Cref{cond:donsker_repr,cond:valid_CB_f}, the studentised statistic \(T_n=\sup_{\x\in\X}|Z_n(\x)|/\widehat\sigma(\x)\) converges weakly to \(T:=\sup_{\x\in\X}|Z(\x)|/\sigma(\x)\). Since \(t_{n,1-\alpha}\) consistently estimates \(Q_T(1-\alpha)\), and since \Cref{cond:valid_CB_f} implies continuity of the distribution of \(T\) at that quantile, 
\[
\Prob\big(|T_n-t_{n,1-\alpha}|\leq \eta_n\big)\to 0.
\]
Therefore,
\[
    \indicatorBig{ (\widehat{T}_n^L-T_n) \wedge 0 \, \leq \, (t_{n,1-\alpha} - T_n) \, < \, (\widehat{T}_n^L-T_n) \vee 0 } = o_p(1).
\]

\textbf{Part (b):} Consider the second summand of \eqref{eqn:decomp_indicat}: \( \displaystyle \indicatorBig{ (t_{n,1-\alpha}^L - t_{n,1-\alpha})\wedge 0 \, < \, (\widehat{T}_n^L - t_{n,1-\alpha}) \, \leq \, (t_{n,1-\alpha}^L - t_{n,1-\alpha}) \vee 0 } \). Under \Cref{cond:valid_CB_f}, the grid-based critical value is controlled by the interpolation error of the studentised bootstrap process.

For a generic bootstrap draw \(Z_n^*\), define the studentised bootstrap process
\[
W_n^*(\x)
:=
\frac{Z_n^*(\x)}{\widehat{\sigma}(\x)},
\]
and let
\[
T_n^L
=
\max_{\x\in\X^L}
|W_n^*(\x)|.
\]
The corresponding infeasible continuous-space statistic is
\[
T_{n,\infty}
:=
\sup_{\x\in\X}
|W_n^*(\x)|.
\]
With the quantile notation in the main text,
\[
t_{n,1-\alpha}^L-t_{n,1-\alpha}
=
\widehat Q_{T_n^L}(1-\alpha)
-
\widehat Q_{T_{n,\infty}}(1-\alpha).
\]

To compare \(T_n^L\) with \(T_{n,\infty}\), introduce the intermediate $d$-linear interpolant of the studentised bootstrap process:
\[
W_{n,L}^*(\x)
:=
\sum_{\bm{\iota}\in\{0,1\}^d}
W_n^*\big(\xbf_L(\x;\bm{\iota})\big)
w_L(\x;\bm{\iota}).
\]
Since \(W_{n,L}^*(\x)\) is multilinear on each interpolation cell and is a convex combination of the corresponding vertex values, its extrema over each cell are attained at the vertices. Consequently,
\[
\sup_{\x\in\X}
|W_{n,L}^*(\x)|
=
\max_{\x\in\X^L}
|W_n^*(\x)|
=
T_n^L.
\]
Therefore,
\begin{align*}
\big|T_n^L-T_{n,\infty}\big|
&=
\left|
\sup_{\x\in\X}
|W_{n,L}^*(\x)|
-
\sup_{\x\in\X}
|W_n^*(\x)|
\right|\\
&\leq
\sup_{\x\in\X}
\left|
W_{n,L}^*(\x)
-
W_n^*(\x)
\right|,
\end{align*}
since the $\sup$ operator is $1$-Lipschitz. Indeed, since the interpolation weights sum to one and each vertex \(\xbf_L(\x;\bm{\iota})\) lies in the same cell as \(\x\),
\begin{align*}
\sup_{\x\in\X}
\left|
W_{n,L}^{*}(\x)-W_n^{*}(\x)
\right|
&\leq
\sup_{\dmet(\x,\y)\leq \delta_{\varepsilon_n}}
\left|
W_n^{*}(\x)-W_n^{*}(\y)
\right|.
\end{align*}
Under \Cref{cond:valid_CB_f}, \(W_n^*\rightsquigarrow_{\widehat{\Prob}}W\) in \(\ell^\infty(\X)\). Conditional weak convergence in \(\ell^\infty(\X)\) implies conditional stochastic equicontinuity in probability. Since \(\delta_{\varepsilon_n}\to0\) under the rate condition $L_n=\omega(r_n^{1/4})$, it follows that
\[
\sup_{\dmet(\x,\y)\leq \delta_{\varepsilon_n}}
\left|
W_n^{*}(\x)-W_n^{*}(\y)
\right|
=o_{\widehat{\Prob}}(1)
\quad\text{in probability}.
\]
Thus,
\[
\sup_{\x\in\X}
\left|
W_{n,L}^{*}(\x)-W_n^{*}(\x)
\right|
=o_{\widehat{\Prob}}(1)
\quad\text{in probability}.
\]

Recall \(t_{n,1-\alpha}^L:=\widehat Q_{T_n^L}(1-\alpha)\) and \(t_{n,1-\alpha}:=\widehat Q_{T_{n,\infty}}(1-\alpha)\). Since
\[
T_n^L
\leq
T_{n,\infty}
+
\sup_{\x\in\X}
\left|
W_{n,L}^*(\x)-W_n^*(\x)
\right|,
\]
and similarly with the roles reversed, the discrepancy between the feasible grid-based critical value and its infeasible continuous-space counterpart is controlled by the uniform interpolation error of the studentised bootstrap process. Thus,
\[
\big|T_n^L-T_{n,\infty}\big|
=o_{\widehat{\Prob}}(1)
\quad\text{in probability}.
\]
This implies that the conditional distribution functions of \(T_n^L\) and \(T_{n,\infty}\) are asymptotically equivalent at continuity points of their common limiting distribution. Since \(T_{n,\infty}\rightsquigarrow_{\widehat{\Prob}}T\), and since the quantile-continuity condition invoked above holds at \(Q_T(1-\alpha)\), the corresponding conditional quantiles satisfy
\[
t_{n,1-\alpha}^L-t_{n,1-\alpha}=o_p(1)
\]
under the rate condition $L_n=\omega(r_n^{1/4})$. Together with \(\widehat T_n^L-T_n=o_p(1)\) from Part (a), the second summand of \eqref{eqn:decomp_indicat} \( \displaystyle \indicatorBig{ (t_{n,1-\alpha}^L - t_{n,1-\alpha})\wedge 0 \, < \, (\widehat{T}_n^L - t_{n,1-\alpha}) \, \leq \, (t_{n,1-\alpha}^L - t_{n,1-\alpha}) \vee 0 } \) is bounded above by
\[
\indicatorBig{|T_n-t_{n,1-\alpha}|
\leq
|\widehat T_n^L-T_n|+|t_{n,1-\alpha}^L-t_{n,1-\alpha}|}.
\]
The right-hand side has a \(o_p(1)\) random radius. Hence, there exists a deterministic sequence \(\eta_n\downarrow 0\) such that
\[
\Prob\big(|\widehat T_n^L-T_n|+|t_{n,1-\alpha}^L-t_{n,1-\alpha}|>\eta_n\big)\to 0.
\]
As in Part (a), continuity of the limiting distribution of \(T\) at its \((1-\alpha)\)-quantile, together with consistency of \(t_{n,1-\alpha}\), implies
\[
\Prob\big(|T_n-t_{n,1-\alpha}|\leq \eta_n\big)\to 0.
\]
Therefore,
\[
\indicatorBig{ (t_{n,1-\alpha}^L - t_{n,1-\alpha})\wedge 0 \, < \, (\widehat{T}_n^L - t_{n,1-\alpha}) \, \leq \, (t_{n,1-\alpha}^L - t_{n,1-\alpha}) \vee 0 } = o_p(1)
\]
under the rate condition $L_n=\omega(r_n^{1/4})$.

It follows from the above, the expectation inequality, and bounded convergence for bounded random variables that converge to zero in probability, that under the rate condition $ L_n=\omega(r_n^{1/4}) $,
\begin{equation}\label{eqn:diff_probs}
    \begin{split}
    \Big|\Prob\Big[F(\x) \in \mathcal{C}_{n,1-\alpha}^L(\x) \ &\forall \ \x\in\X \Big] - \Prob\Big[F(\x) \in \widetilde{\mathcal{C}}_{n,1-\alpha}(\x) \ \forall \ \x\in\X \Big] \Big| \\ 
    &= \Big|\E\Big[ \indicatorBig{ \sup_{\x\in\X} \frac{|\widehat{Z}_n^L(\x)|}{\widehat{\sigma}_L(\x)} \leq t_{n,1-\alpha}^L } - \indicatorBig{ \sup_{\x\in\X} \frac{|Z_n(\x)|}{\widehat{\sigma}(\x)} \leq t_{n,1-\alpha} } \Big] \Big| \\
    &\leq \E\Big[\Big|\indicatorBig{ \sup_{\x\in\X} \frac{|\widehat{Z}_n^L(\x)|}{\widehat{\sigma}_L(\x)} \leq t_{n,1-\alpha}^L } - \indicatorBig{ \sup_{\x\in\X} \frac{|Z_n(\x)|}{\widehat{\sigma}(\x)} \leq t_{n,1-\alpha}} \Big| \Big] \\
    &= o(1).
    \end{split}
\end{equation}
By \Cref{cond:valid_CB_f}, the conclusion follows from \citet[Supplementary Appendix, Lemma SA.1(b)]{chernozhukov-fernandez-melly-2013inference}.

\qed

\section{Additional Monte Carlo Details}\label{app:mc}

This appendix records implementation details not repeated in the Monte Carlo discussion in \Cref{subsec:mc_evidence}. 

\subsection{Monte Carlo Set-up}

In the univariate design, the DTT evaluation region is constructed from the population central interval of the equally weighted mixture of the four group-time outcome distributions. With \(\tau_{\mathrm{lb}}=0.05\) and \(\tau_{\mathrm{ub}}=0.95\),
\[
\underline y_{\mathrm{DTT},1}
=
Q_Y^{\mathrm{mix}}(\tau_{\mathrm{lb}}),
\qquad
\bar y_{\mathrm{DTT},1}
=
Q_Y^{\mathrm{mix}}(\tau_{\mathrm{ub}}),
\]
where \(Q_Y^{\mathrm{mix}}\) is the quantile function of this population mixture. The univariate counterfactual CDF uses \(Q_Y^{\mathrm{CF}}(0.05)\) and \(Q_Y^{\mathrm{CF}}(0.95)\), where \(Q_Y^{\mathrm{CF}}\) is the quantile function of the known population counterfactual distribution.

In the bivariate design, the grid for object \(m\) is a tensor product with \(L_{m,n}^2\) nodes. The evaluation regions are chosen with common coordinate-wise widths, so \(S_m=\bar y_{m,1}-\underline y_{m,1}\) is also the width used along the second coordinate. For DTT, the lower and upper bounds are chosen symmetrically as \(\underline{\y}_{\mathrm{DTT}}=(-c,-c)\) and \(\bar{\y}_{\mathrm{DTT}}=(c,c)\), where \(c>0\) is determined under the equally weighted mixture of the four group-time outcome distributions so that
\[
F(\bar{\y}_{\mathrm{DTT}})-F(\underline{\y}_{\mathrm{DTT}})=0.9.
\]
For the counterfactual CDF, the lower and upper diagonal endpoints \(\underline{\y}_{\mathrm{CF}}\) and \(\bar{\y}_{\mathrm{CF}}\) are chosen so that the known counterfactual lower-orthant probabilities equal \(0.05\) and \(0.95\), respectively. Because the grid rules depend on the object-specific span \(S_m\), the DTT and counterfactual CDF grids may have different realised values of \(L_{m,n}\), as reported in \Cref{Tab:mc_univariate,Tab:mc_bivariate}.

The simulations use the ordinary non-parametric bootstrap. At the level of the first-order representation in \Cref{cond:valid_CB_f}, this resampling scheme corresponds to an exchangeably weighted bootstrap with centred multinomial frequency weights. For the DTT object in the baseline designs, since \(F^{\mathrm{DTT}}\equiv0\), noncoverage is equivalently recorded by the rejection indicator
\[
\varphi_{\mathrm{DTT},n}^{L}
=
\indicatorBig{
\exists\,\x\in\X_{\mathrm{DTT}}
\text{ such that }
0\notin\mathcal{C}_{\mathrm{DTT},n,1-\alpha}^L(\x)
}.
\]

\subsection{Smooth stress design}\label{app:mc_stress}

As a diagnostic check on the fixed-grid mechanism, we also consider a univariate stress design with equally sized group-time cells. The three untreated or pre-treatment cells have \(N(0,1)\) outcomes, while the treated-post cell follows a smooth narrow-bump mixture \((1-\pi_{\mathrm{bump}})N(0,1)+\pi_{\mathrm{bump}}N(0.25,0.12^2)\), with \(\pi_{\mathrm{bump}}=0.08\). The resulting DTT target is smooth but non-zero, and the added local curvature makes coarse interpolation less forgiving. The DTT evaluation region is \([-1.635,1.635]\), while the counterfactual region is \([-1.645,1.645]\).

\begin{table}[!htbp]
\caption{\label{Tab:mc_stress} Stress-test Monte Carlo evidence, \(\R\)-indexed process}
\centering
\scriptsize
\setlength{\tabcolsep}{3pt}
\renewcommand{\arraystretch}{0.92}
\resizebox{\textwidth}{!}{%
\begin{tabular}{@{}clrrrrrrrrrr@{}}
\toprule
& & \multicolumn{5}{c}{DTT} & \multicolumn{5}{c}{CF} \\
\cmidrule(lr){3-7}\cmidrule(lr){8-12}
\(n\) & Rule & \(L\) & \(\mathcal L_2\) & \(90\%\) & \(95\%\) & \(99\%\) & \(L\) & \(\mathcal L_2\) & \(90\%\) & \(95\%\) & \(99\%\) \\
\midrule
\multirow{6}{*}{500} & Fixed & 5 & 0.059 & 0.908 & 0.946 & 0.990 & 5 & 0.051 & 0.877 & 0.950 & 0.989 \\
& Theory & 9 & 0.063 & 0.899 & 0.953 & 0.990 & 9 & 0.054 & 0.892 & 0.949 & 0.989 \\
& \((1,0.30)\) & 22 & 0.065 & 0.895 & 0.951 & 0.992 & 22 & 0.056 & 0.893 & 0.946 & 0.989 \\
& \((2,0.30)\) & 43 & 0.066 & 0.904 & 0.951 & 0.989 & 43 & 0.057 & 0.898 & 0.948 & 0.991 \\
& \((1,0.35)\) & 29 & 0.066 & 0.900 & 0.951 & 0.993 & 29 & 0.057 & 0.898 & 0.949 & 0.992 \\
& \((2,0.35)\) & 58 & 0.067 & 0.901 & 0.945 & 0.989 & 58 & 0.057 & 0.893 & 0.942 & 0.990 \\
\midrule
\multirow{6}{*}{1000} & Fixed & 5 & 0.042 & 0.890 & 0.941 & 0.988 & 5 & 0.038 & 0.868 & 0.937 & 0.991 \\
& Theory & 11 & 0.045 & 0.903 & 0.951 & 0.990 & 11 & 0.039 & 0.894 & 0.952 & 0.990 \\
& \((1,0.30)\) & 26 & 0.046 & 0.902 & 0.954 & 0.991 & 27 & 0.040 & 0.910 & 0.957 & 0.994 \\
& \((2,0.30)\) & 52 & 0.047 & 0.909 & 0.953 & 0.992 & 53 & 0.041 & 0.901 & 0.958 & 0.993 \\
& \((1,0.35)\) & 37 & 0.047 & 0.903 & 0.953 & 0.991 & 37 & 0.040 & 0.907 & 0.954 & 0.992 \\
& \((2,0.35)\) & 74 & 0.047 & 0.905 & 0.954 & 0.991 & 74 & 0.041 & 0.901 & 0.953 & 0.993 \\
\midrule
\multirow{6}{*}{2500} & Fixed & 5 & 0.027 & 0.878 & 0.930 & 0.977 & 5 & 0.025 & 0.826 & 0.898 & 0.978 \\
& Theory & 13 & 0.028 & 0.899 & 0.940 & 0.993 & 13 & 0.024 & 0.905 & 0.953 & 0.989 \\
& \((1,0.30)\) & 35 & 0.029 & 0.899 & 0.942 & 0.993 & 35 & 0.025 & 0.901 & 0.952 & 0.993 \\
& \((2,0.30)\) & 69 & 0.029 & 0.896 & 0.951 & 0.994 & 69 & 0.025 & 0.907 & 0.951 & 0.992 \\
& \((1,0.35)\) & 51 & 0.029 & 0.898 & 0.950 & 0.991 & 51 & 0.025 & 0.904 & 0.950 & 0.992 \\
& \((2,0.35)\) & 102 & 0.029 & 0.899 & 0.947 & 0.991 & 102 & 0.026 & 0.902 & 0.944 & 0.990 \\
\midrule
\multirow{6}{*}{5000} & Fixed & 5 & 0.020 & 0.824 & 0.910 & 0.982 & 5 & 0.020 & 0.704 & 0.815 & 0.948 \\
& Theory & 16 & 0.020 & 0.917 & 0.952 & 0.990 & 16 & 0.017 & 0.904 & 0.946 & 0.985 \\
& \((1,0.30)\) & 43 & 0.020 & 0.908 & 0.952 & 0.994 & 43 & 0.018 & 0.903 & 0.943 & 0.986 \\
& \((2,0.30)\) & 85 & 0.021 & 0.907 & 0.947 & 0.992 & 85 & 0.018 & 0.904 & 0.947 & 0.986 \\
& \((1,0.35)\) & 65 & 0.021 & 0.909 & 0.946 & 0.989 & 65 & 0.018 & 0.896 & 0.946 & 0.987 \\
& \((2,0.35)\) & 129 & 0.021 & 0.912 & 0.945 & 0.989 & 130 & 0.018 & 0.907 & 0.938 & 0.989 \\
\bottomrule
\end{tabular}}
\begin{justify}
\footnotesize Notes: DTT denotes the distributional treatment effect on the treated and CF denotes the counterfactual distribution. The fixed rule sets \(L=5\) for every sample size; Theory uses the \(\kappa=1\) rule; \((a,b)\) indexes the power rule.
\end{justify}
\end{table}

\Cref{Tab:mc_stress} makes the fixed-grid mechanism more transparent than the baseline designs. At small sample sizes, the fixed rule can still look adequate: for the counterfactual CDF, \(95\%\) coverage is \(0.950\) at \(n=500\). As the sample size increases, however, the same fixed grid increasingly under-covers, with counterfactual \(95\%\) coverage falling to \(0.898\) at \(n=2500\) and \(0.815\) at \(n=5000\). The increasing-grid rules remain close to nominal over the same range; for example, the theory-guided rule gives counterfactual \(95\%\) coverage \(0.949\), \(0.952\), \(0.953\), and \(0.946\). The DTT results show the same pattern more mildly, with fixed-rule \(95\%\) coverage declining from \(0.946\) to \(0.910\), while the increasing-grid rules remain near nominal. Thus the stress design supports the asymptotic message without changing the main-text conclusion: fixed grids may look serviceable in moderate samples, but they do not eliminate interpolation error as sampling uncertainty shrinks.

\end{document}